\documentclass{notnws} % nws

\usepackage[usenames,dvipsnames]{xcolor}
\usepackage{amssymb,wasysym,graphicx,hyperref,lipsum,xspace,soul}
\usepackage[textwidth=2.55cm,shadow,disable]{todonotes}

\DeclareGraphicsExtensions{.pdf,.eps,.png,.jpg}

\definecolor{blue}{RGB}{74,128,156}
\definecolor{green}{RGB}{155,187,90}
\definecolor{orange}{RGB}{232,153,42}
\definecolor{gray}{RGB}{153,153,153}

\newcommand{\cmp}[1]{\mathcal{O}(#1)}
\newcommand{\prob}[1]{\mathrm{P}(#1)}
\newcommand{\conv}[1]{\mathcal{H}(#1)}
\newcommand{\neigh}[1]{\Gamma_{#1}}
\newcommand{\ball}[2]{\neigh{#1}(#2)}
\newcommand{\avg}[1]{\langle#1\rangle}
\newcommand{\rnd}[1]{\widetilde{#1}}
\newcommand{\cmd}[1]{\overline{#1}}
\newcommand{\set}[1]{\{\,#1\,\}}

\newcommand{\given}[1]{#1\vert\:}

\newcommand{\dfn}[1]{{\color{green}#1}\xspace}
\newcommand{\dfa}[1]{{\color{blue}#1}\xspace}
\newcommand{\dft}[1]{{\color{gray}#1}\xspace}

\newcommand{\secref}[1]{Section~\ref{sec:#1}\xspace}
\newcommand{\secsref}[2]{Sections~\ref{sec:#1}--\ref{sec:#2}\xspace}
\newcommand{\figref}[1]{Figure~\ref{fig:#1}\xspace}
\newcommand{\figsref}[2]{Figures~\ref{fig:#1}--\ref{fig:#2}\xspace}
\newcommand{\tblref}[1]{Table~\ref{tbl:#1}\xspace}
\newcommand{\eqref}[1]{Equation~(\ref{eq:#1})\xspace}
\newcommand{\appref}[1]{Appendix~\ref{app:#1}\xspace}

\sethlcolor{white}

\newcommand{\revc}[3]{\todo[fancyline,color=\if#11green\else blue\fi!66]{{\color{Black}{\bf Reviewer \##1:}\\#2.\ comment}}{\hl{#3}}\xspace}
\newcommand{\reva}[2]{\todo[fancyline,color=gray!66]{{\color{Black}{\bf Additional:}\\#1.\ change}}{\hl{#2}}\xspace}

\newcommand{\euros}{European highways\xspace}
\newcommand{\celeg}{\emph{Caenorhabditis elegans}\xspace}
\newcommand{\power}{Western US power grid\xspace}
\newcommand{\oreg}{Oregon Internet map\xspace}
\newcommand{\cites}{Scientometrics citations\xspace}

\newcommand{\blogs}{US election weblogs\xspace}
\newcommand{\fweb}{Little Rock food web\xspace}
\newcommand{\flights}{US airports connections\xspace}
\newcommand{\collabs}{Networks coauthorships\xspace}

\newcommand{\ces}{\emph{C.\ elegans}\xspace}

\begin{document}

% % % % % % % % % % % % % % % % % % % % % % % % % % 
%
%				TITLE & AUTHORS
%
% % % % % % % % % % % % % % % % % % % % % % % % % %

\title[Convexity in complex networks]{Convexity in complex networks}

 \author[T. Marc and L. \v{S}ubelj]{
{\sf TILEN MARC}\\\emph{Institute of Mathematics, Physics and Mechanics, Ljubljana, Slovenia}\\\vspace{1em}
{\sf LOVRO \v{S}UBELJ}\\\emph{University of Ljubljana, Faculty of Computer and Information Science, Ljubljana, Slovenia}
\email{lovro.subelj@fri.uni-lj.si}}

\maketitle

% % % % % % % % % % % % % % % % % % % % % % % % % % 
%
%				ABSTRACT
%
% % % % % % % % % % % % % % % % % % % % % % % % % %

\begin{abstract}

Metric graph properties lie in the heart of the analysis of complex networks, while in this paper we study their convexity through mathematical definition of a convex subgraph. A subgraph is convex if every geodesic path between the nodes of the subgraph lies entirely within the subgraph. % We analyze convexity by expanding randomly grown subsets of nodes to convex subgraphs and by observing the frequency of small convex subgraphs. 
According to our perception of convexity, convex network is such in which every connected subset of nodes induces a convex subgraph. We show that convexity is an inherent property of many networks that is not present in a random graph. Most convex are spatial infrastructure networks and social collaboration graphs due to their tree-like or clique-like structure, whereas the food web is the only network studied that is truly non-convex. Core-periphery networks are regionally convex as they can be divided into a non-convex core surrounded by a convex periphery. Random graphs, however, are only locally convex meaning that any connected subgraph of size smaller than the average geodesic distance between the nodes is almost certainly convex. We present different measures of network convexity and discuss its applications in the study of networks.

\paragraph{Keywords:} \emph{network convexity, convex subsets, convex subgraphs, core-periphery structure}

\end{abstract}

% % % % % % % % % % % % % % % % % % % % % % % % % % 
%
%				INTRODUCTION
%
% % % % % % % % % % % % % % % % % % % % % % % % % %

\section{\label{sec:intro}Introduction}

Metric graph theory is a study of geometric properties of graphs based on a notion of the shortest or geodesic path between the nodes defined as the path through the smallest number of edges~\cite{BC08a}. % Metric graph properties lie in the heart of the analysis of complex networks. Classical examples include Milgram's experiment of degrees of separation~\cite{Mil67}, node index called betweenness centrality~\cite{Fre77} and the small-world network model~\cite{WS98}.
\revc{1}{1./3}{Metric graph properties have proved very useful in the study of complex networks in the past}~\cite{Mil67,Fre77,WS98}. Independently of these efforts, metric graph theorists have been interested in understanding convexity in a given graph~\cite{HN81,FJ86,Van93,Pel13}\revc{2}{1}{}. Consider a simple connected graph and a subgraph on some subset of nodes $S$. The subgraph is \emph{induced} if all edges between the nodes in $S$ in the graph are also included in the subgraph. Next, the subgraph is said to be \emph{isometric} if at least one geodesic path joining each two nodes in $S$ is entirely included within $S$. Finally, the subgraph is a \emph{convex subgraph} if all geodesic paths between the nodes in $S$ are entirely included within $S$. For instance, every complete subgraph or a clique is obviously a convex subgraph. Notice that any convex subgraph is also isometric, while any isometric subgraph must necessarily be induced.

\begin{figure}
	\includegraphics[width=0.95\textwidth]{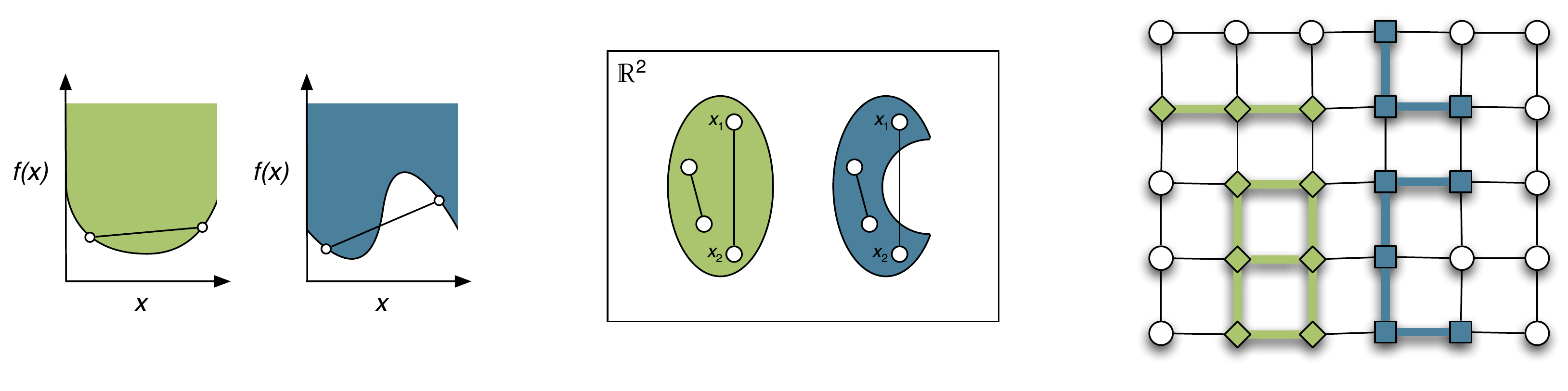}%
	\caption{\label{fig:convex}Standard definitions of convexity for different mathematical objects. (\emph{left})~Real-valued function $f(x)$ is convex if the line segment between any two points $(x_1,f(x_1))$ and $(x_2,f(x_2))$ is above or on the graph of $f$, $\forall t\in[0,1]\colon\: tf(x_1)+(1-t)f(x_2)\geq f(tx_1+(1-t)x_2)$. (\emph{middle})~Set $S\subset\mathbb{R}^2$ is convex if the line segment between any two points $x_1,x_2\in S$ lies entirely within $S$, $\forall t\in[0,1]\colon\: tx_1+(1-t)x_2\in S$. (\emph{right})~Connected subgraph induced by a subset of nodes $S$ is convex if any geodesic path between two nodes in $S$ goes exclusively through $S$ (\dfn{diamonds}). Otherwise, the subgraph is non-convex (\dfa{squares}).}
\end{figure}

For better understanding, \figref{convex} compares standard definitions of convexity for %a real-valued function, a set in an Euclidean plane and a subgraph of a simple graph. 
\revc{1}{1}{different mathematical objects}. In all %these 
cases, convexity of a mathematical object is defined through the inclusion of the shortest or geodesic paths between its parts.

\begin{figure}
	\includegraphics[width=0.6\textwidth]{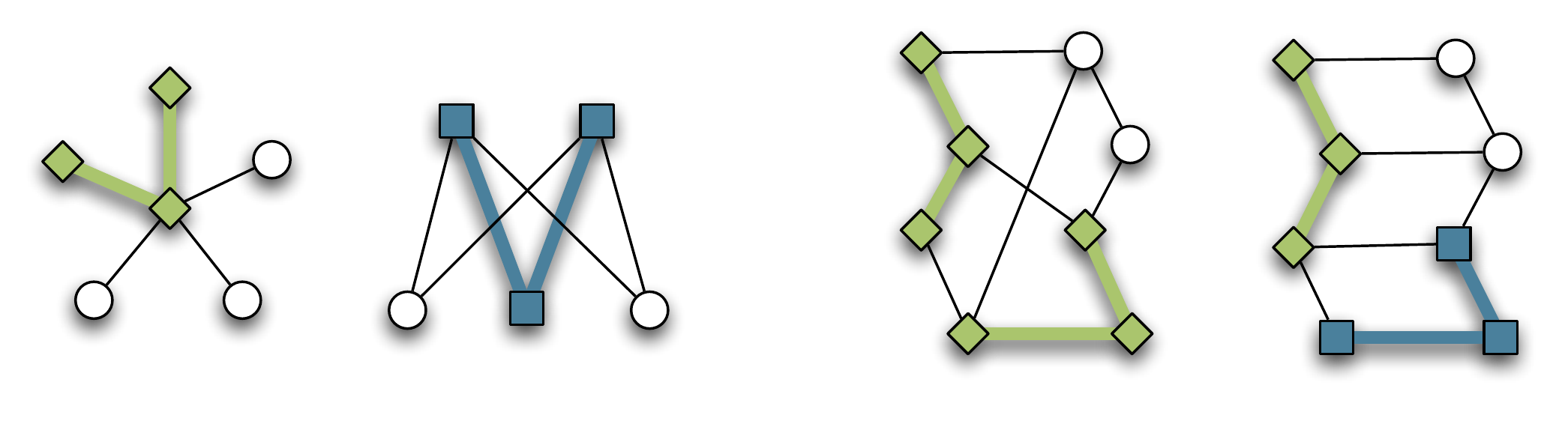}%
	\caption{\label{fig:examps}Pairs of different graphs with the same or similar number of induced subgraphs, but varying numbers of convex (\dfn{diamonds}) and non-convex (\dfa{squares}) subgraphs. For instance, all connected triples of nodes are convex subgraphs in the first graph of each pair.}
\end{figure}

Convex subgraphs provide an insight into the metric structure of graphs as building blocks for embedding them in simple metric spaces~\cite{Van93,BC08a,Pel13}\revc{2}{1}{}. See the two graphs shown in the left side of~\figref{examps}. The first one is a star graph representing hub-and-spokes arrangement found in airline transportation networks~\cite{Bar11b} and the Internet~\cite{GSA07a}. The second one is a bipartite graph suitable for modeling two-mode affiliation networks~\cite{DGG41} or word adjacency networks~\cite{MIKLSASA04}. From the perspective of either graph theory or network science, these two graphs would be deemed different. However, they both contain no triangles and $10$ or $9$ connected triples of nodes, which is quite similar. On the other hand, all connected triples of nodes in the first graph are convex subgraphs (\dfn{diamonds}), whereas none is convex in the second graph (\dfa{squares}). In this way, convex subgraphs are very sensitive to how they are intertwined with the rest of the graph.

One probably noticed that the two graphs differ in the number of nodes and edges. The right side of~\figref{examps} therefore shows two additional graphs that are identical up to $3$-node subgraphs.\revc{1}{1}{} %They both contain $8$ nodes, $10$ edges, $16$ connected triples of nodes and no triangles. 
Yet, the graphs are obviously different. Looking at their convex subgraphs again nicely discriminates between the two as all subgraphs in the first graph are convex.

Convex subgraphs explore convexity in graphs only locally. Define the \emph{convex hull} $\conv{S}$ of a subset of nodes $S$ to be the smallest convex subgraph including $S$~\cite{HN81}. Since the intersection of convex subgraphs is also a convex subgraph, $\conv{S}$ is uniquely defined. Now the \emph{hull number} of a graph is the size of the smallest $S$ whose $\conv{S}$ is the entire graph~\cite{ES85}. This number can be interpreted as a convexity-based measure exploring the global macroscopic structure of a graph. For instance, the hull numbers of the two graphs in the left side of~\figref{examps} are $5$ and $2$, while computing the hull number of a general graph is NP-hard~\cite{DGKPS09}.

The concept of convexity is by no means novel to the study of networks. Social networks literature defines a clique to be a maximal group of \revc{1}{4}{nodes} directly connected by \hl{an edge}. As such definition might be too crude for larger groups, a $k$-clique is defined as a group of \hl{nodes} at distance at most $k$~\cite{Luc50}. For $k=1$, one recovers the original definition of a clique. Finally, a $k$-clan further restricts that all geodesic paths must lie within the group~\cite{WF94}, which is precisely our understanding of convexity. Still, there is no restriction on the maximum distance $k$ in the definition of a convex subgraph. The nodes can be \revc{1}{5}{at} any distance as long as the subgraph is convex.

% Community detection has been one of the most popular research topics in network science~\cite{FH16}. Informally, a community is a densely connected group of nodes that is only loosely connected to the rest of the network. Edge betweenness method defines communities as connected groups obtained after removing the edges supporting the largest number of geodesic paths~\cite{GN02}. An opposite approach is taken by the map equation method~\cite{RB08}, where communities are revealed directly by means of a trapped random walk. Clique percolation method detects overlapping communities by merging together adjacent cliques~\cite{PDFV05}. In all~these cases, the adopted definition of community can be seen as an approximation of a convex~subgraph.

\revc{1}{1./6}{}The analysis of small subgraphs or fragments~\cite{Bat88,EK15} in empirical networks is else known under different terms. Motifs refer to not necessarily induced subgraphs whose frequency is greater than in an appropriate random graph model~\cite{MSIKCA02}. Graphlets, however, are induced subgraphs that represent specific local patterns found in biological and other networks~\cite{PCJ04}. Small subgraphs have proven extremely useful in network comparison~\cite{MIKLSASA04,Prv07a} and, recently, for uncovering higher-order connectivity in networks~\cite{XWC16,BGL16}. Note that some of the subgraphs are convex by construction or very (un)likely to be convex under any random graph model. In this sense, the above work already provides a glimpse of convexity in complex networks.

% In this paper we study convexity in more general terms by asking ``What is convexity \emph{in} complex networks?''. (Similarly as a subset of a plane can be convex or not, while a plane is always convex, a subgraph can be convex or not, whereas a connected graph would always be convex. Thus, asking ``What is convexity \emph{of} complex networks?'' would make little sense.) In~\secref{expans}, we first study convexity from a global macroscopic perspective by analyzing the expansion of convex subsets of nodes in graphs and networks. In~\secref{freqs}, we support our findings by analyzing convexity also locally through the frequency of small convex subgraphs. \secref{measures} discusses various forms of convexity observed in graphs and networks, and proposes different measures of convexity. \secref{concs} concludes the paper with the discussion of network convexity and future work.

In this paper we study convexity in more general terms by asking ``What is convexity \emph{in} complex networks?''. (Similarly as a subset of a plane can be convex or not, while a plane is always convex, a subgraph can be convex or not, whereas a connected graph would always be convex. Thus, asking ``What is convexity \emph{of} complex networks?'' would make little sense.) \revc{1}{1./7}{We try to answer this question by expanding randomly grown subsets of nodes to convex subgraphs and observing their growth, and by comparing the frequency of small convex subgraphs to non-convex subgraphs. This allows us to study convexity from a global macroscopic perspective while also locally.

We demonstrate several distinct forms of convexity in graphs and networks. Networks characterized by a tree-like or clique-like structure are globally convex meaning that any connected subset of nodes will likely induce a convex subgraph. This is in contrast with random graphs that are merely locally convex meaning that only subgraphs of size smaller than the average geodesic distance between the nodes are convex. Core-periphery networks are found to be regionally convex as they can be divided into a non-convex core surrounded by a convex periphery. Convexity is thus an inherent structural property of many networks that is not present in a random graph. It can be seen as an indication of uniqueness of geodesic paths in a network, which in fact unifies the structure of tree-like and clique-like networks. This property is neither captured by standard network measures nor is convexity reproduced by standard network models. We therefore propose different measures of convexity and argue for its use in the future studies of networks.}

\hl{The rest of the paper is structured as follows.} In~\secref{expans}, we first study convexity from a global perspective by analyzing the expansion of convex subsets of nodes. In~\secref{freqs}, we support our findings by analyzing convexity also locally through the frequency of small convex subgraphs. \secref{measures} discusses various forms of convexity observed in graphs and networks, and proposes different measures of convexity. \secref{concs} concludes the paper with the discussion of network convexity and \hl{prominent directions for} future work.

% % % % % % % % % % % % % % % % % % % % % % % % % % 
%
%				EXPANSIONS
%
% % % % % % % % % % % % % % % % % % % % % % % % % %

\section{\label{sec:expans}Expansion of convex subsets of nodes}

We study convexity in different regular and random graphs, synthetic networks and nine empirical networks from various domains. These represent power supply lines of the western US power grid~\cite{WS98}, highways between European cities part of the E-road network in 2010~\cite{SB11d}, coauthorships between network scientists parsed from the bibliographies of two review papers in 2006~\cite{New06a}, Internet map at the level of autonomous systems reconstructed from the University of Oregon Route Views Project in 2000~\cite{LKF07}, protein-protein interactions of the nematode \celeg collected from the BioGRID repository in 2016~\cite{SBRBBT06}, connections between US airports compiled from the Bureau of Transportation Statistics data in 2010~\cite{Kun13}, citations between scientometrics papers published in \emph{Journal of Informetrics}, \emph{Scientometrics} or \emph{JASIST} between 2009-2013 as in the Web of Science database~\cite{SVW16a}, hyperlinks between weblogs on the US presidential election of 2004~\cite{AG05} and predator-prey relationships between the species of Little Rock Lake~\cite{WM00}. 

\begin{table}%
	\caption{\label{tbl:nets}\emph{Basic statistics of empirical networks studied in the paper. These show the number of nodes $n$ and edges $m$, the average node degree $\avg{k}$ and clustering coefficient $\avg{C}$, and the average geodesic distance between the nodes $\avg{\ell}$.}}
	\begin{tabular}{lrrrrr} \hline\hline
		Network & $n$ & $m$ & $\avg{k}$ & $\avg{C}$ & $\avg{\ell}$ \\\hline
		\power & $4941$ & $6594$ & $2.67$ & $0.08$ & $18.99$ \\
		\euros & $1039$ & $1305$ & $2.51$ & $0.02$ & $18.40$ \\
		\collabs & $379$ & $914$ & $4.82$ & $0.74$ & $6.04$ \\
		\oreg & $767$ & $1734$ & $4.52$ & $0.29$ & $3.03$ \\
		\celeg & $3747$ & $7762$ & $4.14$ & $0.06$ & $4.32$ \\
		\flights & $1572$ & $17214$ & $21.90$ & $0.50$ & $3.12$ \\
		\cites & $1878$ & $5412$ & $5.76$ & $0.13$ & $5.52$ \\
		\blogs & $1222$ & $16714$ & $27.36$ & $0.32$ & $2.74$ \\
		\fweb & $183$ & $2434$ & $26.60$ & $0.32$ & $2.15$ \\\hline\hline
	\end{tabular} 
\end{table}

The networks are listed in~\tblref{nets}. Although some of the networks are directed, all are represented with simple undirected graphs and reduced to the largest connected component. \tblref{nets} also shows the basic statistics of the networks including the number of nodes $n$ and edges $m$, the average node degree $\avg{k}$, $\avg{k}=2m/n$, the average node clustering coefficient $\avg{C}$~\cite{WS98} with the clustering coefficient of node $i$ defined as $C_i = \frac{2t_i}{k_i(k_i-1)}$, where $k_i$ is the degree and $t_i$ is the number of triangles including node $i$, and the average geodesic distance between the nodes $\avg{\ell}$, $\avg{\ell}=\frac{1}{n}\sum_i\ell_i$, where $\ell_i=\frac{1}{n-1}\sum_{j\neq i}d_{ij}$ and $d_{ij}$ is the geodesic distance between the nodes $i$ and $j$ defined as the number of edges in the geodesic path. The networks are ordered roughly by decreasing average geodesic distance $\avg{\ell}$, which will become clear later on.

Given a particular network or graph, we define a subset of nodes $S$ to be a \emph{convex subset} when the subgraph induced by $S$ is a convex subgraph. In what follows, we study convexity by analyzing the growth of convex subsets of nodes and observing how fast they expand. Recall the hull number defined as the size of the smallest subset $S$ whose convex hull $\conv{S}$ spans the entire network~\cite{ES85}. Since $\conv{S}$ is the smallest convex subgraph including $S$, the hull number measures how \emph{quickly} convex subsets can grow. We here take the opposite stance and analyze how \emph{slowly} randomly grown convex subsets expand. We use an algorithm for expansion of convex subsets, which we present next.

We start by initializing \revc{1}{8}{a} subset $S$ with a randomly selected seed node. We then grow $S$ one node at a time and observe the evolution of its size. To ensure convexity, $S$ is expanded to the nodes of its convex hull $\conv{S}$ on each step. Every $S$ realized by the algorithm is thus a convex subset. Newly added nodes are selected among the neighbors of nodes in $S$ by following a random edge leading outside of $S$. In other words, new nodes are selected with the probability proportional to the number of neighbors they have in $S$. This ensures that $S$ is a slowly growing connected subset of nodes. An alternative approach would be to select new nodes uniformly at random from the neighboring nodes.

Let $\neigh{i}$ denote the set of neighbors of node $i$. The complete algorithm for convex subset expansion is given below. % (see~\appref{expans} for pseudocode and algorithmic complexity).
\begin{enumerate}
	\item Select random seed node $i$ and set $S=\set{i}$.
	\item Until $S$ contains all nodes repeat the following:
	\begin{enumerate}
		\item Select node $i\notin S$ with probability $\propto |\neigh{i}\cap S|$.
		\item Expand $S$ to the nodes of $\conv{S\cup\set{i}}$. 
	\end{enumerate}
\end{enumerate}

Before looking at the results, it is instructive to consider the evolution of $S$ in the first few steps of the algorithm. Initially, $S$ contains a single node $i$, $S=\set{i}$, which is a convex subset. Next, one of its neighbors $j$ is added, $S=\set{i,j}$, which is still convex. On the next step, a neighbor $k$ of say $j$ is added to $S$, $S=\set{i,j,k}$. If $k$ is also a neighbor of $i$, $S$ is a convex subset. This is expected in a network that is locally clique-like indicated by high clustering coefficient $\avg{C}>0.5$. Similarly, in a (locally) tree-like network with zero clustering coefficient $\avg{C}\approx 0$, every connected triple of nodes including $S$ is expected to be convex. In any other case, $S$ would have to be expanded with all common neighbors of $i$ and $k$, which may demand additional nodes and so on, possibly resulting in an abrupt growth of $S$. Therefore, in the early steps of the algorithm, the expansion of convex subsets quantifies the presence of locally tree-like or clique-like structure in a network. In the later steps, the algorithm explores also higher-order connectivity, whether a network is tree-like or clique-like as a whole. In the extreme case of a tree or a complete graph, every connected subset of nodes induces either a tree or a clique, which are both convex subgraphs.

\begin{figure}
	\includegraphics[width=0.8\textwidth]{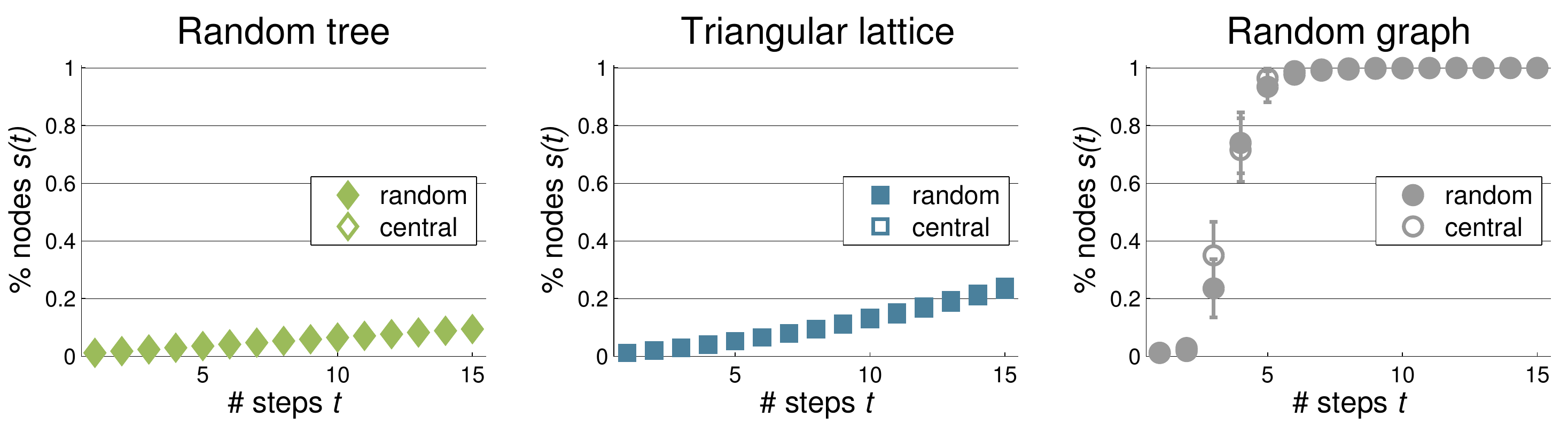}\\\hskip16pt
	\includegraphics[width=0.2667\textwidth]{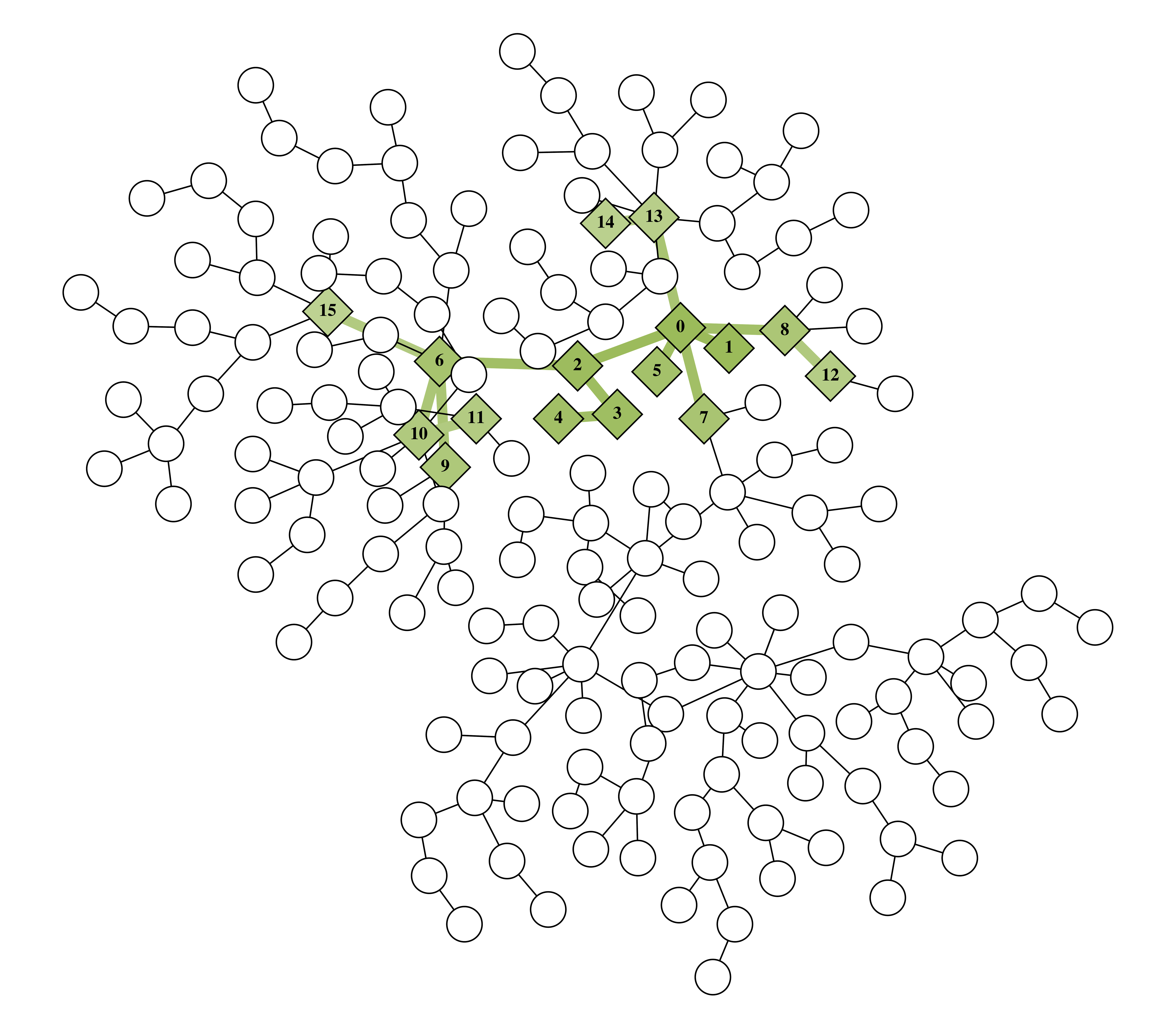}%
	\includegraphics[width=0.2667\textwidth]{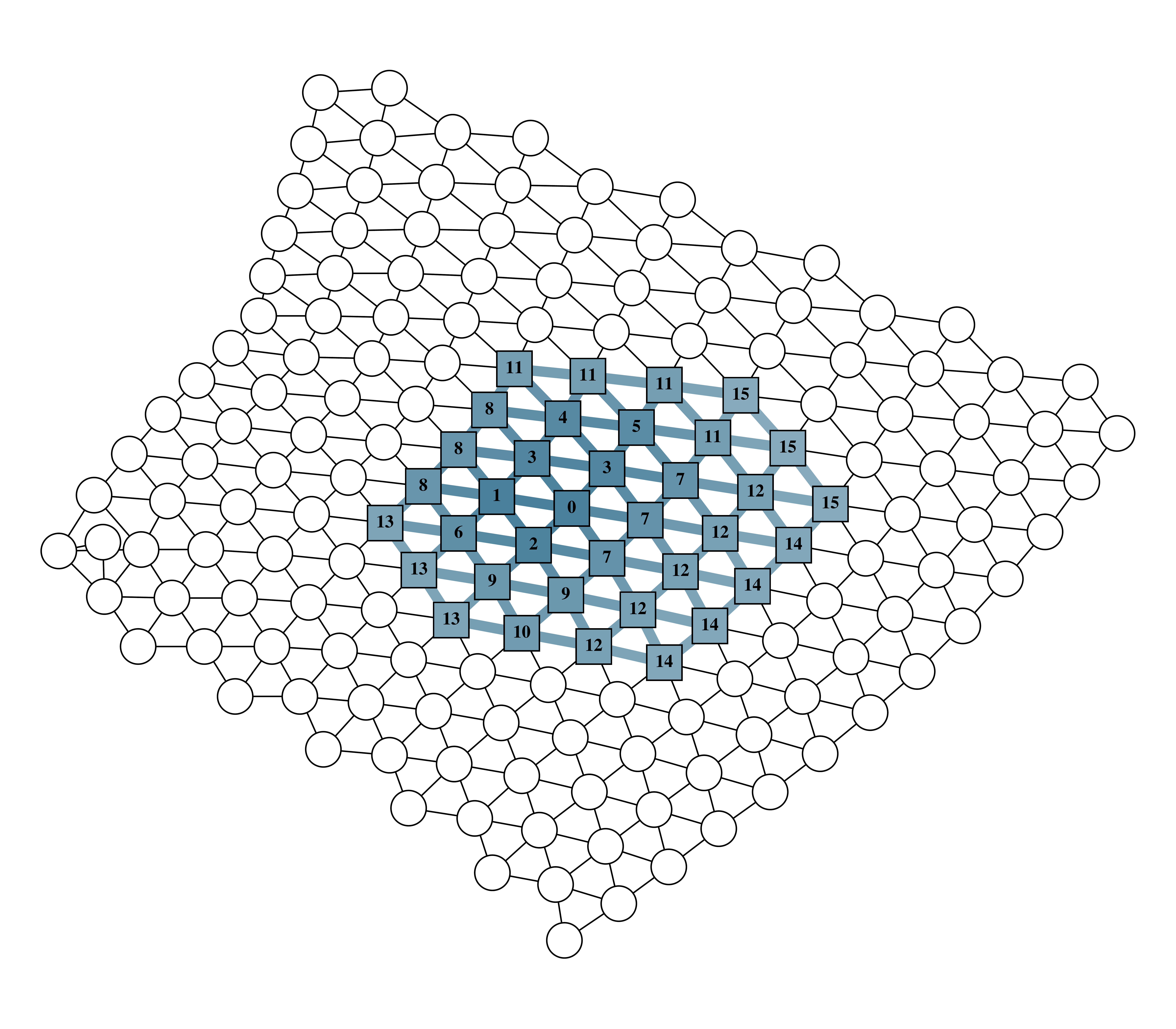}%
	\includegraphics[width=0.2667\textwidth]{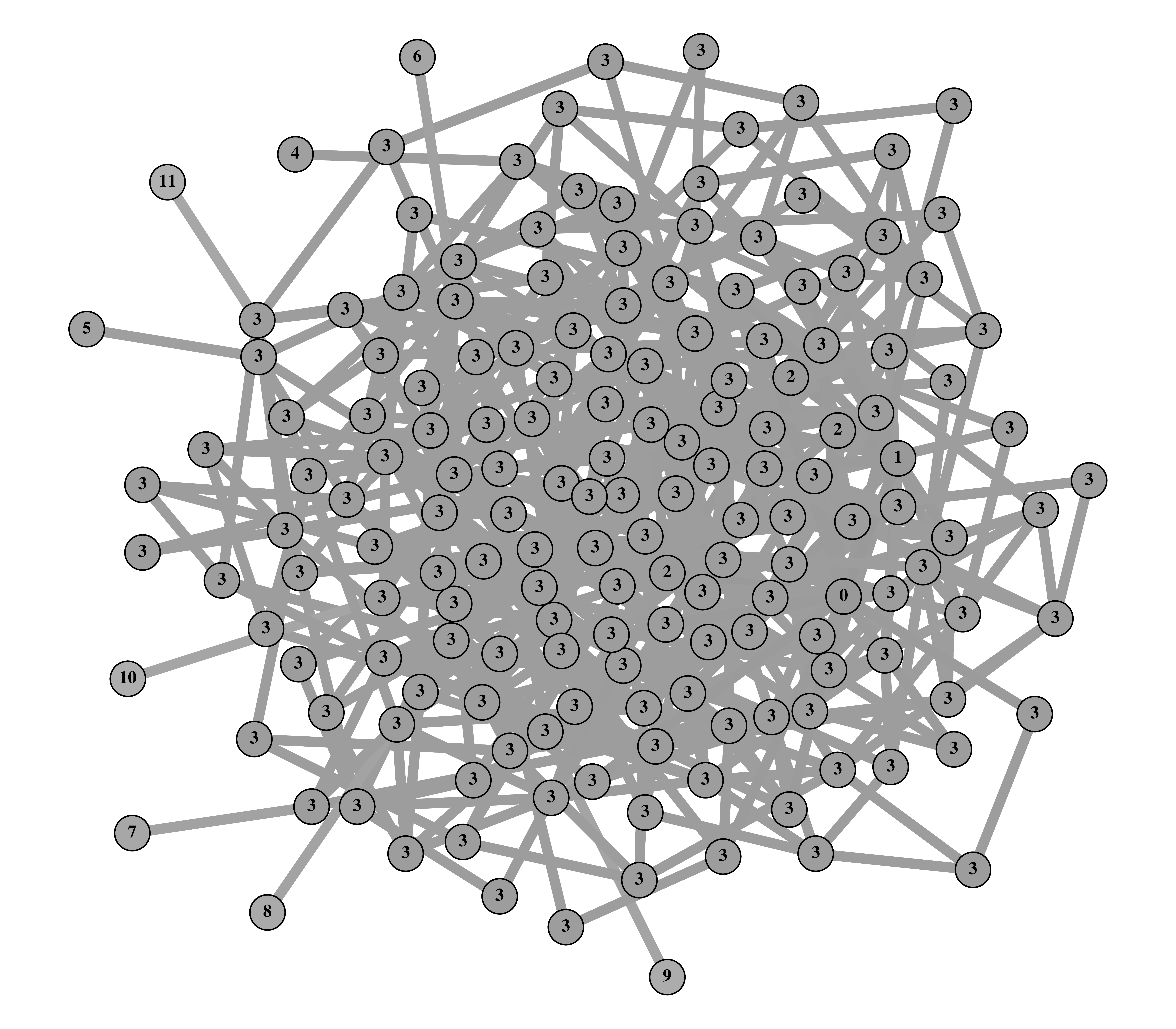}%
	\caption{\label{fig:graphs}Expansion of convex subsets of nodes in a randomly grown tree (\dfn{diamonds}), triangular lattice of the same size (\dfa{squares}) and the corresponding random graph (\dft{ellipses}). (\emph{top})~The fractions of nodes $s(t)$ in the growing convex subsets at different steps $t$ of our algorithm. The subsets are grown from a seed node selected uniformly at random and the most central node with the smallest geodesic distance $\ell$ to other nodes. The markers are estimates of the mean over $100$ runs, while error bars show the $99\%$ confidence intervals. (\emph{bottom})~Highlighted subgraphs show particular realizations of convex subsets grown from the most central node for $15$ steps as in the plots above. The labels of the nodes indicate the step $t$ in which they were included in the convex subset.}
\end{figure}

\figref{graphs} shows the evolution of $S$ in a randomly grown tree on $169$ nodes, triangular lattice with the side of $13$ nodes and a random graph~\cite{ER59} with $169$ nodes and the same number of edges as the lattice. The plots show the fraction of nodes $s(t)$ included in the subset $S$ at different steps $t$ of the algorithm, $t\geq 0$. Note that $t$ is the number of expansion steps (2.\ step), disregarding the initialization step (1.\ step). Hence, $s(0)=1/n$, $s(1)=2/n$ and $s(2)\geq 3/n$. In general, $s(t)\geq (t+1)/n$.

As anticipated above, convex subsets grow one node at a time in a tree graph (\dfn{diamonds} in~\figref{graphs}). Although somewhat counterintuitive, the same slow growth also occurs in a complete graph (results not shown). Relatively modest growth is observed in a triangular lattice due to its nearly clique-like structure (\dfa{squares} in~\figref{graphs}). However, in a random graph (\dft{ellipses} in~\figref{graphs}), convex subsets grow slowly only in the first few steps, due to its locally tree-like structure, upon which they expand rapidly to include all the nodes.

\begin{figure}
	\includegraphics[width=0.8\textwidth]{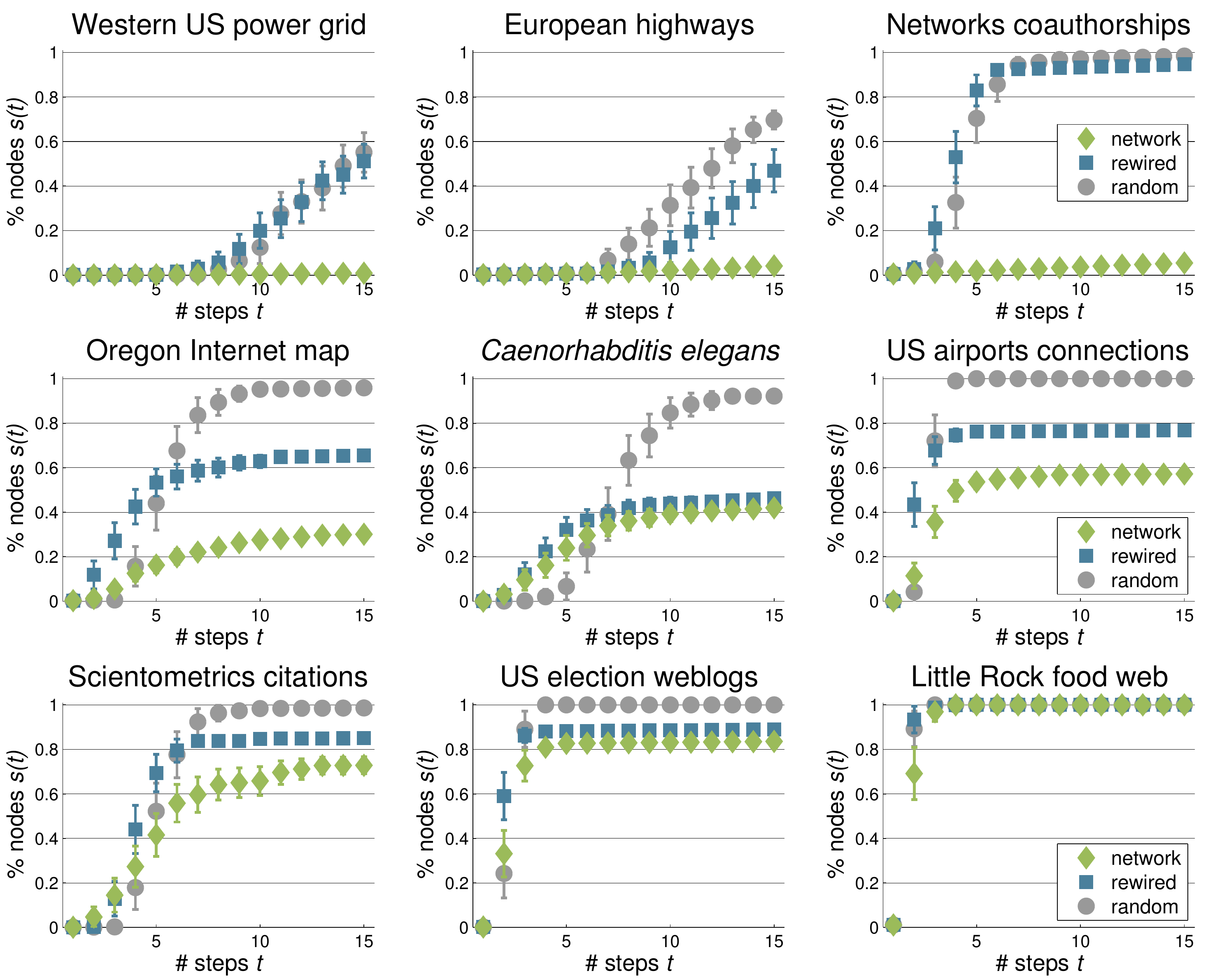}%
	\caption{\label{fig:expans}Expansion of convex subsets of nodes in empirical networks (\dfn{diamonds}), randomly rewired networks % or the configuration model graphs 
	(\dfa{squares}) and the corresponding Erd{\H o}s-R{\'e}nyi random graphs (\dft{ellipses}). Plots show the fractions of nodes $s(t)$ in the growing convex subsets at different steps $t$ of our algorithm. The markers are estimates of the mean over $100$ runs, while error bars show the $99\%$ confidence intervals.}
\end{figure}

\revc{1}{9}{}\figref{expans} shows the evolution of $S$ in empirical networks introduced in~\tblref{nets}, random graphs with the same number of nodes $n$ and edges $m$~\cite{ER59} % and randomly rewired networks by degree preserving randomization. These are random graphs with the same node degree sequence $k_1,k_2,\dots,k_n$ as the original networks more commonly known as the configuration model graphs~\cite{NSW01}. 
\revc{1}{9}{and random graphs with the same node degree sequence $k_1,k_2,\dots,k_n$. These are obtained by randomly rewiring the original networks using $10m$ steps of degree preserving randomization}~\cite{MS02}\revc{1}{10}{}. For relevant comparison, we ensure that all realizations of random graphs are simple and connected.

Notice substantial differences between the networks (\dfn{diamonds} in~\figref{expans}). Convex subsets grow almost one node at a time in the western US power grid and European highways network. These are both spatial infrastructure networks that are locally tree-like with very low clustering coefficient $\avg{C}\approx 0$. The same slow growth is also observed in the coauthorship graph that is locally clique-like with $\avg{C}=0.74$. Note that the lack of any sudden growth indicates that the structure of these networks is throughout tree-like or clique-like. In other networks, convex subsets expand relatively quickly in the early steps, whereas the growth settles after a certain fraction of nodes has been included. This occurs after including $42\%$ of the nodes in \ces protein network, while over $83\%$ in the weblogs graph. Finally, the growth in the food web is almost instantaneous, where convex subsets expand by entire trophic levels and thus cover the network in just a couple of steps.

Erd{\H o}s-R{\'e}nyi random graphs fail to reproduce the trends observed in empirical networks (\dft{ellipses} in~\figref{expans}). In the top row of~\figref{expans}, random graphs match the growth in networks only in the first few steps, due to reasons explained above, whereupon the convex subsets expand quite rapidly. The difference is most pronounced in the case of the coauthorship graph. We consider this an important finding as it shows that convexity is an inherent property of some networks. In contrast, in the bottom rows of~\figref{expans}, convex subsets initially grow faster in networks than in random graphs, while they settle already after including some finite fraction of nodes. Notice, however, that the expansion always occurs at about the same number of steps, which is best observed in the citation network.

% Configuration model graphs 
\revc{1}{9}{Randomly rewired networks} show similar trends as random graphs (\dfa{squares} in~\figref{expans}), yet the convex subsets settle much sooner. In a particular case of \ces protein network, the growth of convex subsets seems to be entirely explained by node degrees.

According to our perception of convexity, a convex network or graph is such in which every connected subset of nodes is convex. Convexity is therefore associated with extremely slow growth of convex subsets as in the spatial infrastructure networks and social coauthorship graph, whereas non-convexity can be identified by instantaneous growth as in the food web. By measuring this growth, one can analyze convexity quantitatively. We return to this in~\secref{measures}, while next, in~\secref{sizes}, we first show that the expansion of convex subsets in networks occurs when the number of steps of our algorithm exceeds the average geodesic distance between the nodes and, in~\secref{cores}, that the growth settles when the convex subsets extend to the network core.

\subsection{\label{sec:sizes}Size of convex subsets of nodes}

Expansion of convex subsets in empirical networks and random graphs occurs at about the same number of steps of our algorithm (see bottom row of~\figref{expans}). Below we show that this happens when the number of steps $t$ of the algorithm exceeds the average geodesic distance $\avg{\ell}$ in a network, $t>\avg{\ell}$, or an appropriate estimate for a regular or random graph.

\begin{figure}
	\includegraphics[width=0.8\textwidth]{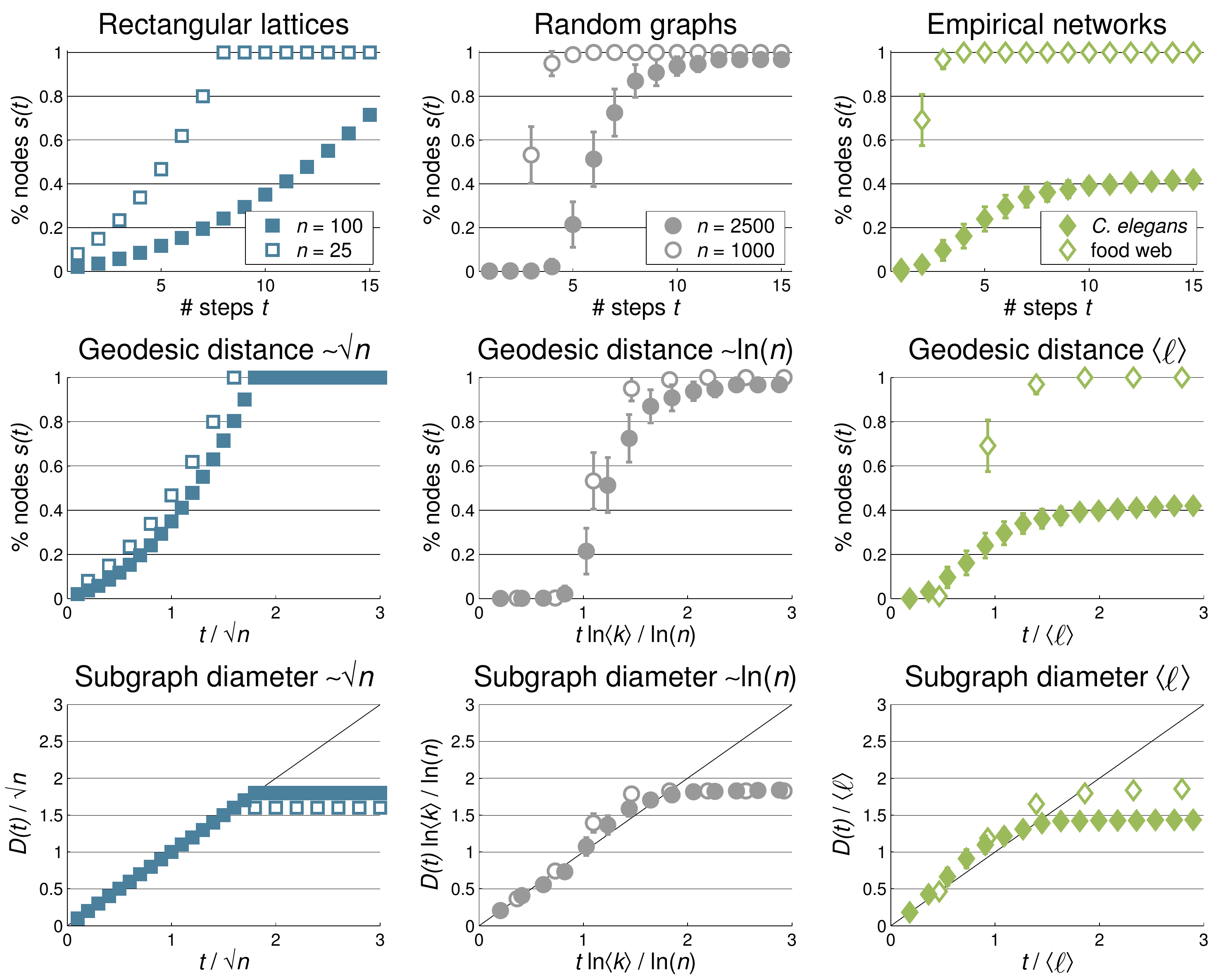}%
	\caption{\label{fig:diams}Expansion of convex subsets of nodes in rectangular lattices with different number of nodes $n$ (\dfa{squares}), Erd{\H o}s-R{\'e}nyi random graphs with different $n$ but the same number of edges $m$ (\dft{ellipses}) and empirical networks with different average geodesic distance $\avg{\ell}$ (\dfn{diamonds}). (\emph{top})~The fractions of nodes $s(t)$ in the growing convex subsets at different steps $t$ of our algorithm. (\emph{middle})~The graphs of $s(t)$ with the steps $t$ rescaled by the average geodesic distance between the nodes. (\emph{bottom})~The growth of the diameter $D(t)$ of convex subgraphs at different steps $t$ shown using the rescaled variables as above. The markers are estimates of the mean over $100$ runs, while error bars show the $99\%$ confidence intervals.}
\end{figure}

Top row of~\figref{diams} shows the evolution of convex subsets $S$ in rectangular lattices with the side of $5$ and $10$ nodes (\dfa{squares}), Erd{\H o}s-R{\'e}nyi random graphs with the number of nodes $n$ equal to $1000$ and $2500$, and the average node degree $\avg{k}$ equal to $12.5$ and $5$, respectively (\dft{ellipses}), and two empirical networks with relatively different average geodesic distance $\avg{\ell}$ (\dfn{diamonds}). These pairs of graphs and networks were specifically selected since they show distinct trends with the expansion occurring at different number of steps~$t$.

\revc{1}{11}{The} middle row of~\figref{diams} shows the evolution of $S$ with the steps $t$ rescaled by the average geodesic distance between the nodes. We use the empirical value $\avg{\ell}$ for networks, and the analytical estimates $\sqrt{n}$ for lattices and $\ln{n}/\ln{\avg{k}}$ for random graphs~\cite{New10}. % Observe that the expansion of convex subsets in graphs and networks takes place when the rescaled number of steps exceeds one. 
\revc{1}{12}{Notice that the expansion of convex subsets in graphs and networks occurs when the rescaled number of steps 
$t\ln{\avg{k}}/\ln{n}$ or $t/\avg{\ell}$, respectively, becomes larger than one.} There is a sudden transition in random graphs at $t=\ln{n}/\ln{\avg{k}}$, while the growth is much more gradual in networks and settles when $\avg{\ell}<t<2\avg{\ell}$. Similar trend is observed also in rectangular lattices. In what follows, we give a probabilistic argument for this behavior relevant for networks and graphs, while we %provide a formal proof 
\revc{1}{2./13}{further formalize the results} for random graphs in~\appref{sizes}.

Consider a pair of nodes at the maximum geodesic distance or diameter in the subgraph induced by $S$. Let $D(t)$ be the diameter of the subgraph at step $t$ and let $d(t)$ be the geodesic distance between the mentioned nodes in the complete network or graph that is internally disjoint from $S$. Note that $S$ is a convex subset only if $d(t)>D(t)$. For $d(t)\leq D(t)$, not all geodesic paths between the mentioned nodes are included in the subgraph and the expansion of $S$ occurs. If $S$ is small enough then the average geodesic distance $\avg{\ell}$ in the network is almost identical as in the remaining network obtained after removing all the nodes in $S$ but the mentioned ones. Now assume that $D(t)\geq \avg{\ell}$. Since the mentioned nodes can be considered arbitrary in the remaining network, simply by the properties of an average $\prob{d(t) \leq D(t)}>0.5$. Hence, when the diameter of the subgraph $D(t)$ exceeds the average geodesic distance $\avg{\ell}$ in a network or an appropriate estimate for a graph, there is a significant probability that $S$ is not a convex subset and that the expansion of $S$~will~occur.

Bottom row of~\figref{diams} shows the evolution of subgraph diameter $D(t)$. Due to small diameter of graphs and networks considered, $D(t)$ initially grows linearly with the number of steps $t$. In the case of rectangular lattices, every convex subset $S$ induces a rectangular sublattice with the side of the sublattice increased by one on each step $t$. It is thus easy to see that $D(t)=t$. In networks and random graphs, $D(t)\approx t$ as long as the number of steps $t$ is below the average geodesic distance $\avg{\ell}$ in a network or $\ln{n}/\ln{\avg{k}}$ in a random graph. However, when $t\geq\avg{\ell}$ or $t\geq\ln{n}/\ln{\avg{k}}$, also $D(t)\geq\avg{\ell}$ or $D(t)\geq\ln{n}/\ln{\avg{k}}$, and by the above argument the expansion of $S$ is expected to occur.

Regardless of this equivalence, the expansion of convex subsets in networks and random graphs is still notably different (see middle row of~\figref{diams}). There is a sudden growth in random graphs at $t=\ln{n}/\ln{\avg{k}}$, whereas every connected subset with up to $\ln{n}/\ln{\avg{k}}$ nodes is almost certainly convex (see \revc{1}{2./13}{derivation in}~\appref{sizes}). We refer to this as \emph{local convexity}. On the other hand, networks in~\figref{diams} are not locally convex with the expansion starting already when $t<\avg{\ell}$. Furthermore, in~\secref{freqs}, we show that even the most convex infrastructure networks and coauthorship graph from the beginning of~\secref{expans} do not match the local convexity of random graphs.

\subsection{\label{sec:cores}Non-convex core and convex periphery}

Expansion of convex subsets in empirical networks settles after including a certain fraction of nodes (see bottom rows of~\figref{expans}). Although every run of our algorithm is of course different, convex subsets actually converge to the same subsets of nodes in these networks. More precisely, for sufficient number of runs of the algorithm, each node is included in either more than $90$-$95\%$ or less than $10$-$15\%$ of the grown convex subsets, with no node in between. Below we analyze the convex subsets grown for $15$ steps as in~\figsref{graphs}{diams} and show that these are in fact the cores of the networks.

\begin{figure}
	\includegraphics[width=0.5\textwidth]{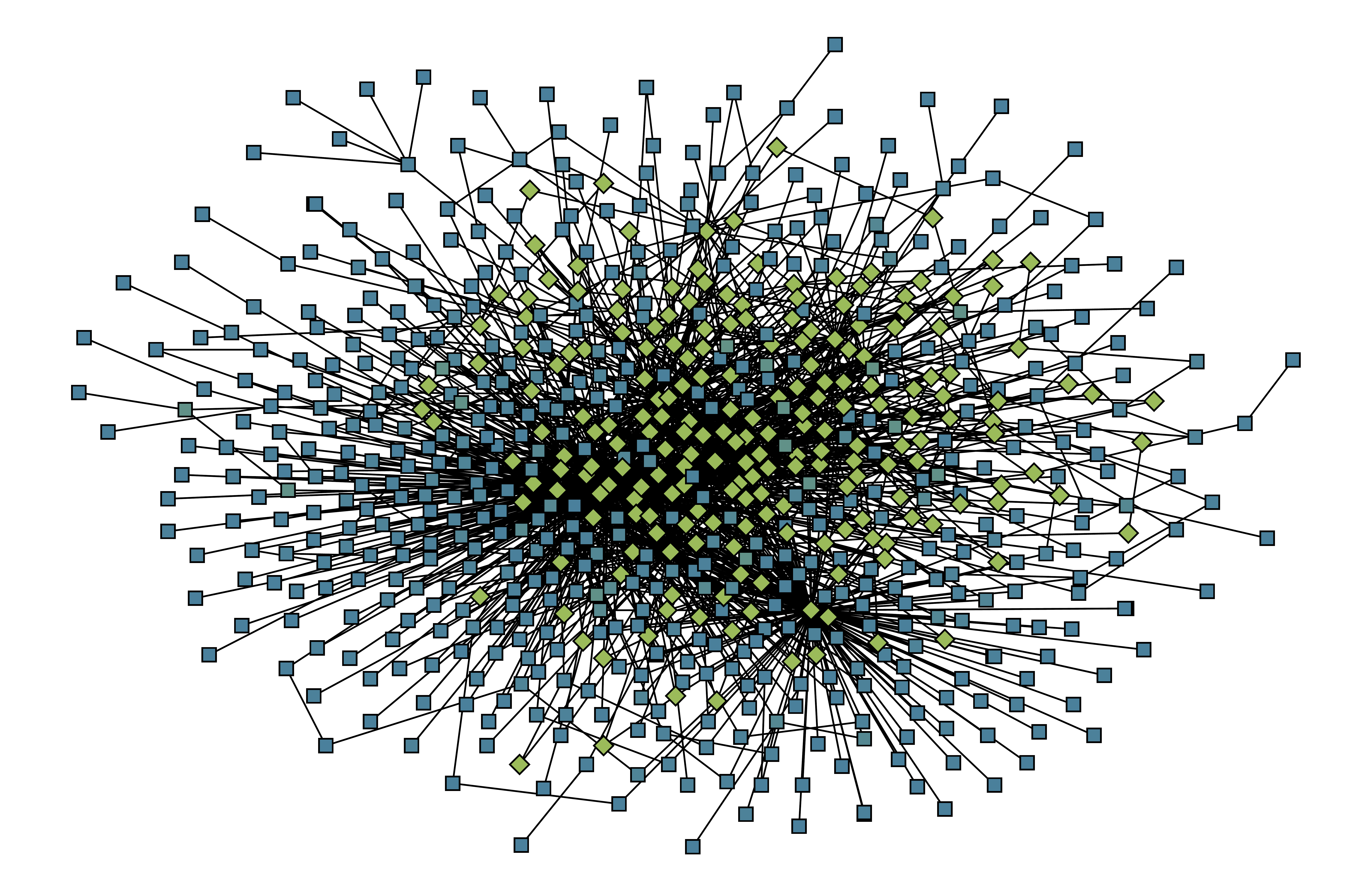}%
	\includegraphics[width=0.5\textwidth]{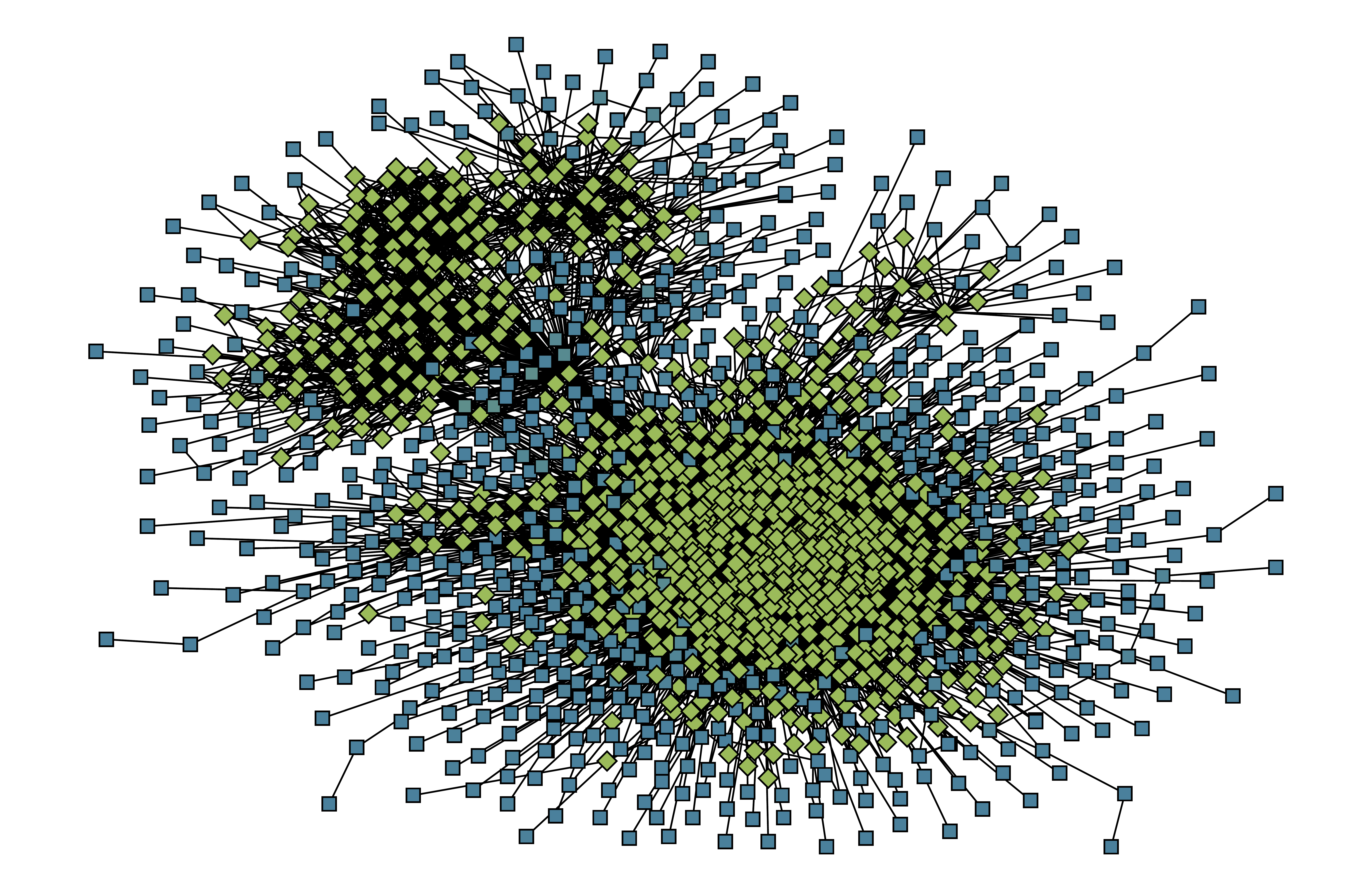}%
	\caption{\label{fig:exmpls}Division of (\emph{left})~the Internet map and (\emph{right})~airline transportation network into core (\dfn{diamonds}) and periphery (\dfa{squares}). The cores are convex subsets grown for $15$ steps of our algorithm. Network layouts were computed with Large Graph Layout~\protect\cite{ADWM04}.}
\end{figure}

Core-periphery structure refers to a natural division of many networks into a densely connected core surrounded by a sparse disconnected periphery~\cite{BE00}. There exist different interpretations of core-periphery structure~\cite{Hol05} including those based on the $k$-core decomposition~\cite{Sei83}, blockmodeling~\cite{DBF05}, stochastic block models~\cite{ZMN15}, conductance cuts~\cite{LLDM09}, overlapping communities~\cite{YL12a} and others~\cite{RPFM14}. Formally, core-periphery structure can be defined by requiring that the probability of connection within the core is larger than between the core and the periphery, which is further larger than within the periphery. For the division into core and periphery inferred from the grown convex subsets in the Internet map and airline transportation network in~\figref{exmpls}, these probabilities are $36.8\permil$, $6.5\permil$, $0.9\permil$ and $40.4\permil$, $2.2\permil$, $0.5\permil$, respectively.

\begin{figure}
	\includegraphics[width=0.8\textwidth]{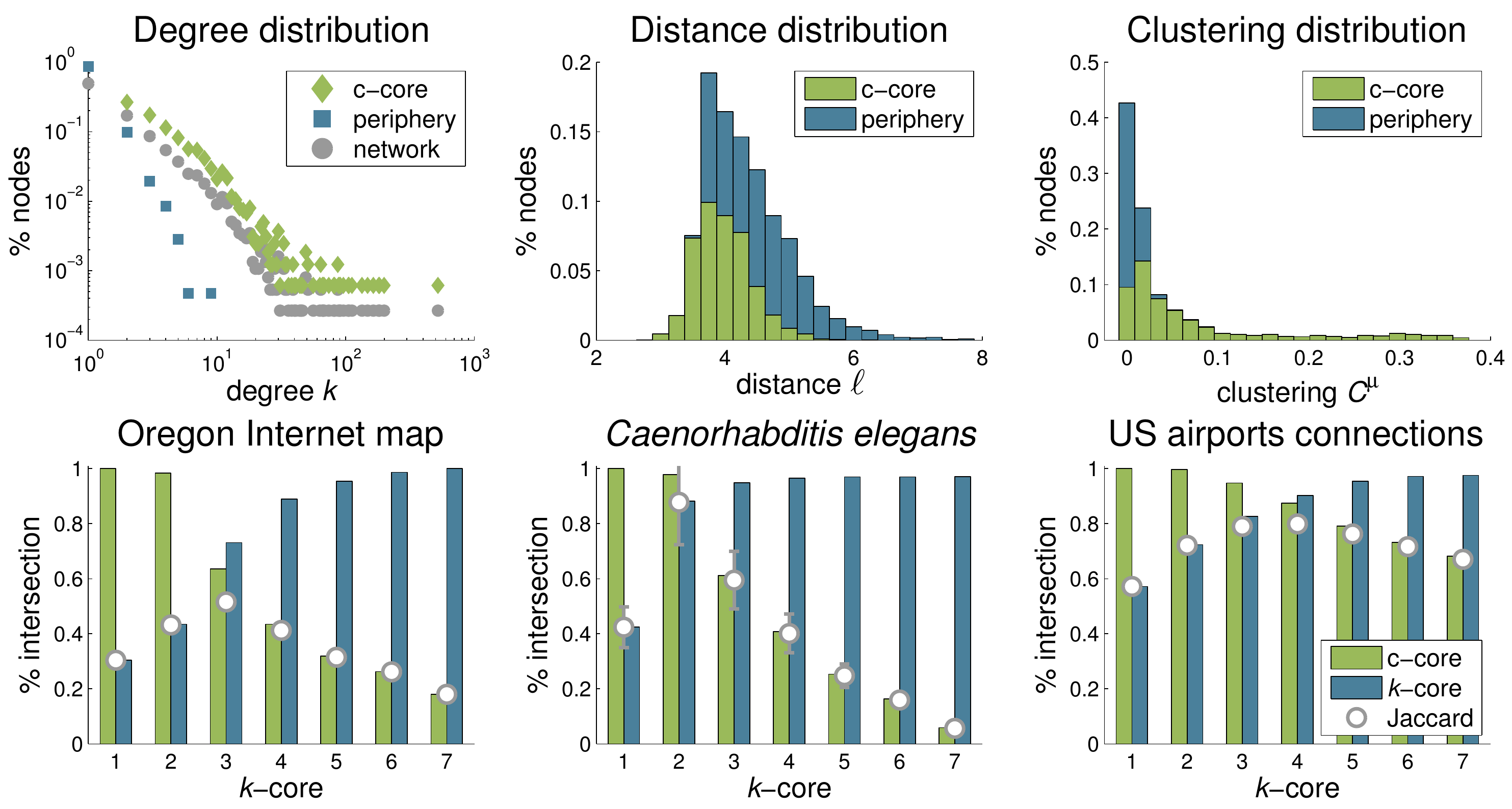}%
	\caption{\label{fig:cores}Analysis of the core-periphery structure identified by our algorithm in empirical networks. (\emph{top})~The distributions of node degree $k$ and geodesic distance $\ell$ in \ces protein network, and the distribution of corrected node clustering coefficient $C^{\mu}$ in airline transportation network, separately for the nodes in the c-core (\dfn{diamonds}) and the periphery (\dfa{squares}). (\emph{bottom})~Comparison between the c-core and the $k$-core decomposition. Plots show the fractions of nodes in the intersection of c-core and $k$-core for different $k$ relative to the size of one or the other, and the Jaccard coefficient of the two subsets. The markers are estimates of the mean over $100$ runs, while error bars show the standard~deviation.}
\end{figure}

Hereafter we refer to the nodes included in at least $90\%$ of the grown convex subsets as the convexity core or c-core for short and to the remaining nodes as the periphery. Top row of~\figref{cores} shows different distributions separately for the nodes in the c-core (\dfn{diamonds}) and the periphery (\dfa{squares}). These are the distributions of node degree $k$ and the average geodesic distance to other nodes $\ell$, $\ell_i=\frac{1}{n-1}\sum_{j\neq i}d_{ij}$, for \ces protein network, and the distribution of corrected node clustering coefficient $C^\mu$~\cite{Bat16}, where $C_i^\mu = \frac{2t_i}{k_i\mu}$ and $\mu$ is the maximum number of triangles a single edge belongs to, for airline transportation network due to low clustering of the former (see~\tblref{nets}). Notice that the nodes in the c-core have higher degrees and also clustering coefficient than peripheral nodes, while they also occupy a more central position in the network with lower geodesic distances to other nodes. Besides, network degree distribution seems to be entirely governed by the c-core, whereas the nodes in the periphery follow a different seemingly scale-free distribution. Although interesting on its own, we do not investigate this further here.

Core-periphery division identified by our algorithm is compared against the $k$-core decomposition~\cite{Sei83} that gained much attention recently~\cite{BDLMG15,YDSH16,HGA16}. A $k$-core is a maximal subset of nodes in which every node is connected to at least $k$ others. It can be identified by iteratively pruning the nodes with degree less than $k$ until no such node remains~\cite{BZ11}. Since every $k$-core is a subset of a $(k-1)$-core and so on, $k$-cores form a nested decomposition of a network, with $1$-core being the set of all nodes in a connected network. Nodes in a $k$-core that are not part of a $(k+1)$-core are called a $k$-shell. Note that $k$-cores can be disconnected, which is not the case for the networks below.

Bottom row of~\figref{cores} shows the fraction of nodes in the intersection of c-core and $k$-core for different $k$ relative to the number of nodes in one subset or the other. In the case of \ces protein network, the c-core is almost entirely included within the $2$-core and contains $88\%$ of the nodes of the $2$-core and $95\%$ of the nodes of the $3$-core. In airline transportation network, the c-core best matches the $4$-core and contains $90\%$ of its nodes, while it also contains nodes from the $3$-shell and $2$-shell. On the other hand, the c-core of the Internet map shows low similarity to any $k$-core. Core-periphery structure identified by our algorithm thus differs from the $k$-core decomposition. % According to our knowledge, this is also the first study of the core-periphery structure based on the inclusion of geodesic paths, although the geodesic paths themselves have already been considered before~\cite{CRLP16}.
\revc{1}{15}{According to our knowledge, this is the first study of the core-periphery structure based on the \emph{inclusion} of geodesic paths, whereas the length and number of geodesic paths has already been considered before}~\cite{Hol05,CRLP16}.

We stress that despite the fact that the c-core of these networks is a convex subset by definition, a c-core is a \emph{non-convex} core surrounded by a \emph{convex} periphery according to our understanding of convexity. This is because convex subsets expand very quickly until they reach the edge of the c-core beyond which the growth settles. In other words, the c-core is the smallest convex subset including the network core. Convexity in core-periphery networks can therefore be interpreted in terms of the size of the c-core. In this sense, convex infrastructure networks from the beginning of~\secref{expans} have no \reva{1}{c-core}, while non-convex food web lacks periphery.

% % % % % % % % % % % % % % % % % % % % % % % % % % 
%
%				FREQUENCIES
%
% % % % % % % % % % % % % % % % % % % % % % % % % %

\section{\label{sec:freqs}Frequency of small convex subgraphs}

\secref{expans} explores convexity in graphs and networks from a global macroscopic perspective, while in this section we analyze convexity also locally. We study small connected induced subgraphs or graphlets in biological networks jargon~\cite{PCJ04,Prv07a} and ask whether induced subgraphs found in empirical networks are convex subgraphs. Note that this is fundamentally different from expanding subsets of nodes to convex subgraphs and observing their growth as in~\secref{expans}. We nevertheless expect networks that have proven extremely convex or non-convex in that global sense to be such also locally.

\begin{figure}
	\includegraphics[width=0.8\textwidth]{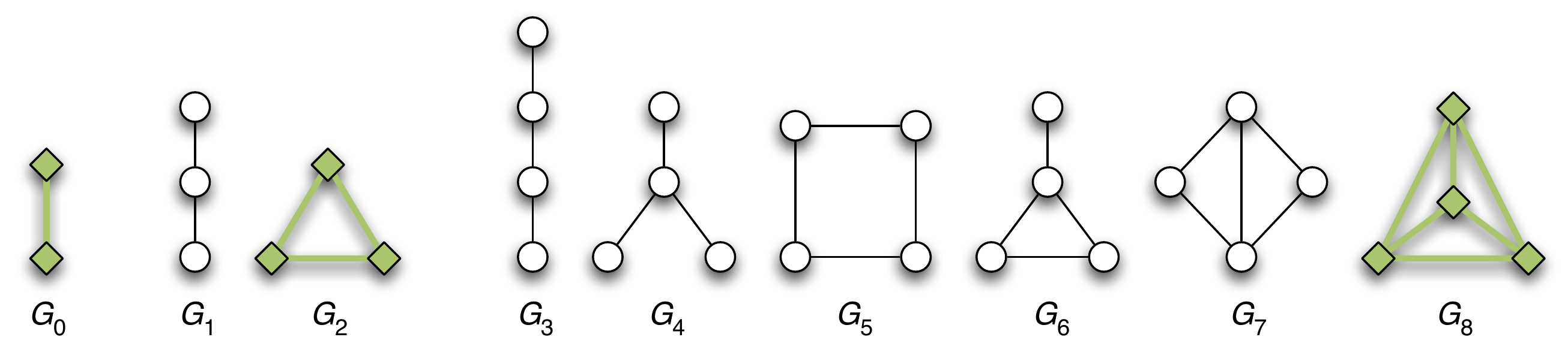}%
	\caption{\label{fig:subgraphs}Connected non-isomorphic subgraphs with up to four nodes. Highlighted subgraphs are convex by construction (\dfn{diamonds}), while the trivial edge subgraph $G_0$ is shown only for reasons of completeness.}
\end{figure}

We consider subgraphs $G_i$ with at most four nodes shown in~\figref{subgraphs}. Note that prior probabilities of convexity vary across subgraphs. The clique subgraphs $G_0$, $G_2$ and $G_8$ are convex by construction (\dfn{diamonds}), whereas the path subgraph $G_3$ is the least likely to be convex. The frequencies of induced subgraphs are computed with a combinatorial method \cite{HD14}, while we use our own implementation for convex~subgraphs.

\begin{figure}
	\includegraphics[width=0.8\textwidth]{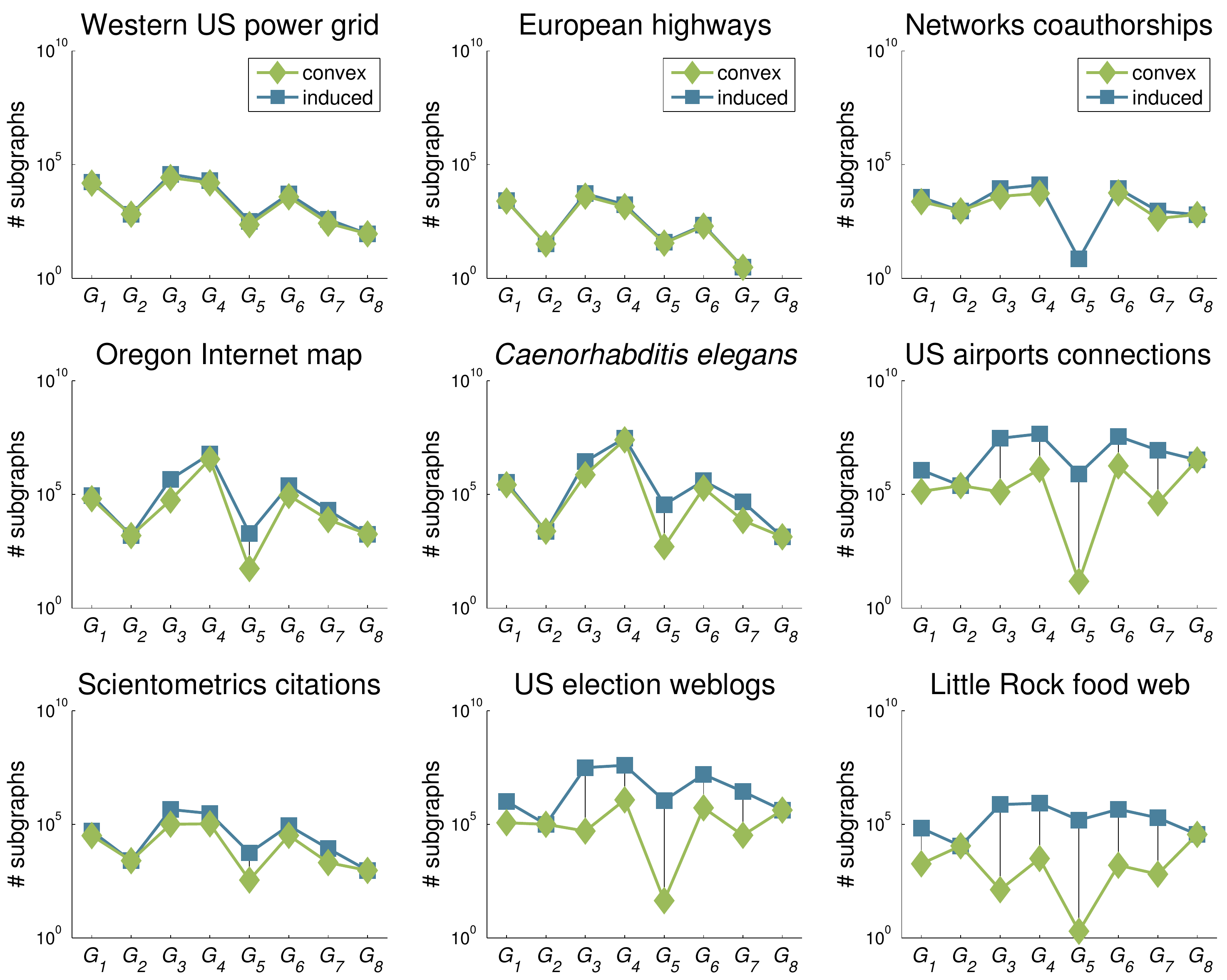}%
	\caption{\label{fig:freqs}Frequency of small non-isomorphic subgraphs $G_i$ in empirical networks. Plots show the number of induced subgraphs $g_i$ (\dfa{squares}) and the number of these that are convex $c_i$ (\dfn{diamonds}). The subgraphs $G_i$ are listed in~\figref{subgraphs}, while the lines are merely a guide for the eye.}
\end{figure}

\figref{freqs} shows the frequencies of induced (\dfa{squares}) and convex (\dfn{diamonds}) subgraphs $G_i$ in networks from~\tblref{nets}. In the case of infrastructure networks, most induced subgraphs are convex subgraphs. Similar holds for the coauthorship graph. On the contrary, only a small fraction of subgraphs is convex in the food web or the weblogs graph (mind logarithmic scales). These are precisely the networks that were identified as either particularly convex or non-convex by the expansion of convex subsets in~\figref{expans}. This confirms that convexity is an inherent property of some networks independent of the specific view taken. There are a few differences relative to before which we~discuss~below.

\begin{figure}
	\includegraphics[width=0.8\textwidth]{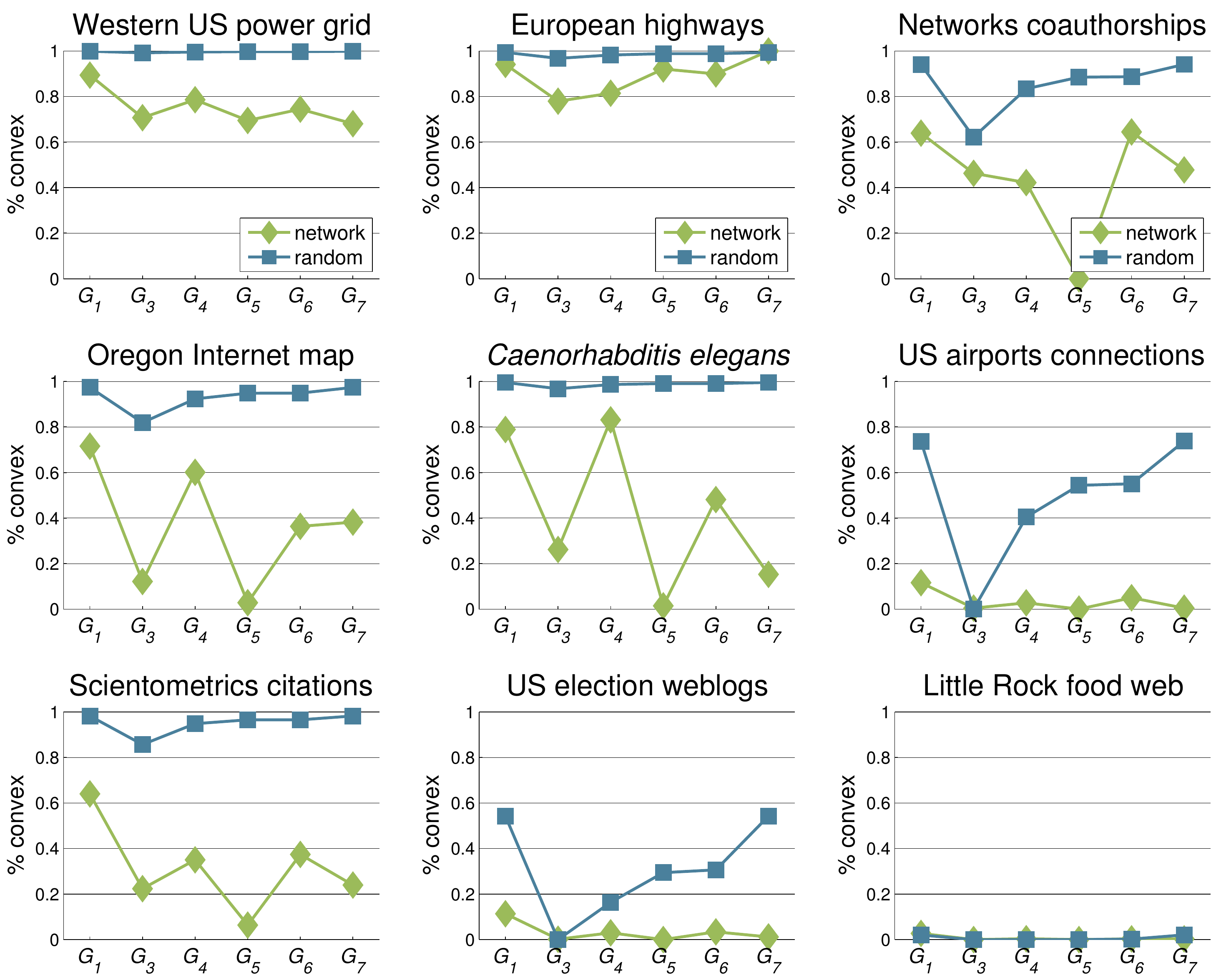}%
	\caption{\label{fig:probs}Probability of convex subgraphs $G_i$ in empirical networks $P_i$ (\dfn{diamonds}) and the corresponding Erd{\H o}s-R{\'e}nyi random graphs $\rnd{P}_i$ (\dfa{squares}). The frequencies of subgraphs $G_i$ are shown in~\figref{freqs}, while the lines are merely a guide for the eye.}
\end{figure}

Let $g_i$ be the number of induced subgraphs $G_i$ in a network and let $c_i$ be the number of these that are convex subgraphs. The empirical probability $P_i$ that a randomly selected subgraph $G_i$ is convex is then
\begin{eqnarray}
	P_i & = & \frac{c_i}{g_i}. \label{eq:Pi}
\end{eqnarray}
In~\appref{probs}, we also derive the analytical priors $\rnd{P}_i$ that a random subgraph $G_i$ is convex in a corresponding Erd{\H o}s-R{\'e}nyi random graph. As already shown in~\secref{sizes}, random graphs are locally convex meaning
that any connected subgraph with up to $\ln{n}/\ln{\avg{k}}$~nodes is expected to be convex. This includes also the subgraphs $G_i$ in all but very dense~graphs.

\figref{probs} shows the empirical probabilities $P_i$ (\dfn{diamonds}) excluding those of clique subgraphs for obvious reasons. As observed above, most subgraphs are convex in the infrastructure networks with $P_i\approx 80\%$, while almost none is convex in the food web or the weblogs graph with $P_i\approx 0\%$. Notice, however, that no square subgraph $G_5$ is convex in the coauthorship graph that was previously classified as convex. Yet, only seven subgraphs $G_5$ appear in the entire network, thus a random subgraph is still more likely to be convex. Non-negligible fractions of subgraphs are convex $P_i>50\%$ also in the Internet map and \ces protein network. Recall that both of these networks have a pronounced core-periphery structure with a relatively small c-core (see~\secref{cores}). The majority of subgraphs is thus found in the periphery which is convex.

\figref{probs} shows also the prior probabilities $\rnd{P}_i$ (\dfa{squares}) that are consistently higher $\rnd{P}_i>P_i$ and tend to $100\%$ in sparse networks. Note that notably lower $\rnd{P}_i$ in airline transportation network, the weblogs graph and the food web are due to much higher density of these networks $\avg{k}>20$ (see~\tblref{nets}). In the case of all other networks, every subgraph $G_i$ is almost certainly convex in a corresponding random graph $\rnd{P}_i\approx 100\%$. Hence, even the most convex infrastructure networks do not match the local convexity of random graphs, despite being considerably more convex from a global point of view.

% % % % % % % % % % % % % % % % % % % % % % % % % % 
%
%				MEASURES
%
% % % % % % % % % % % % % % % % % % % % % % % % % %

\section{\label{sec:measures}Measures of convexity in networks}

\begin{figure}
	\includegraphics[width=0.33\textwidth]{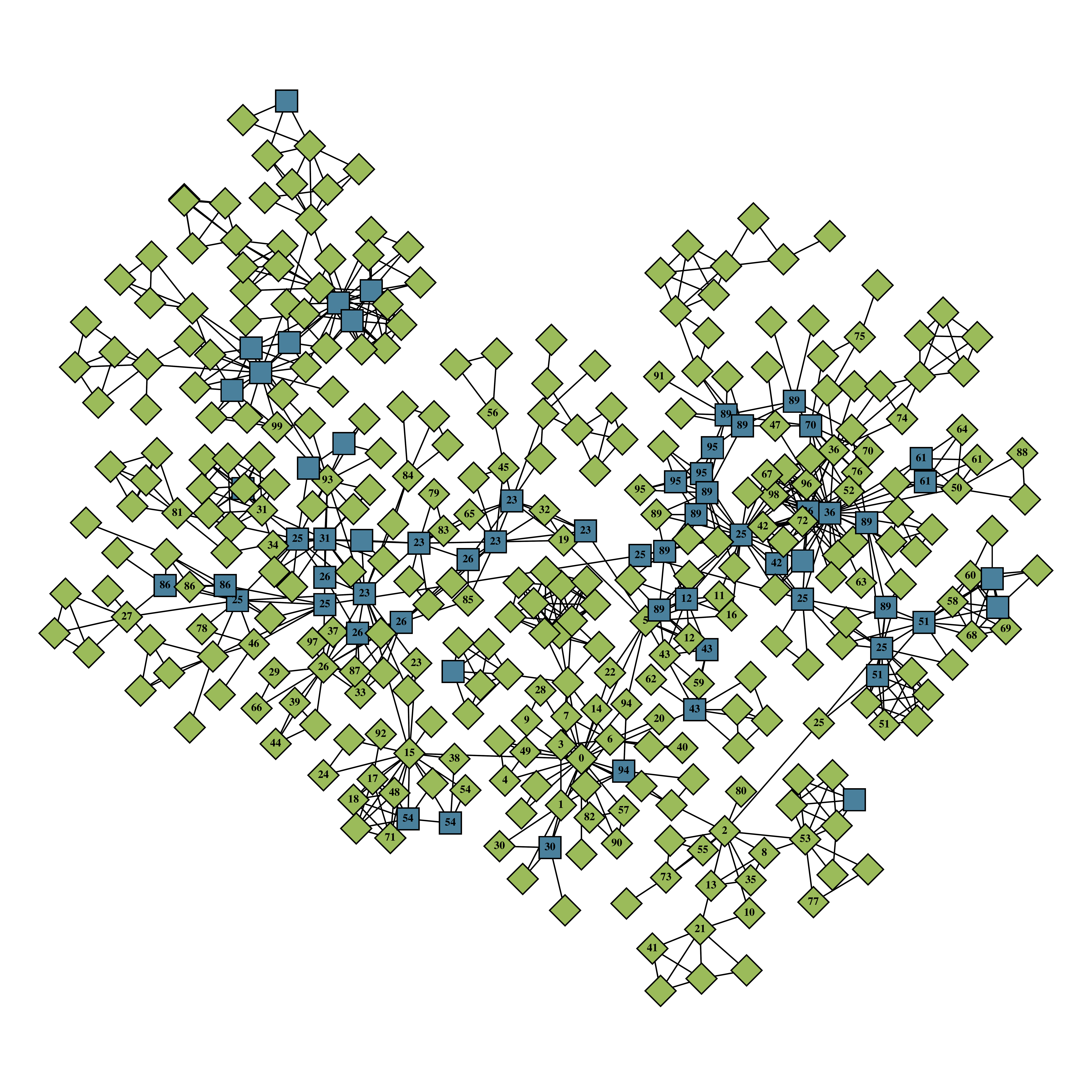}%
	\includegraphics[width=0.33\textwidth]{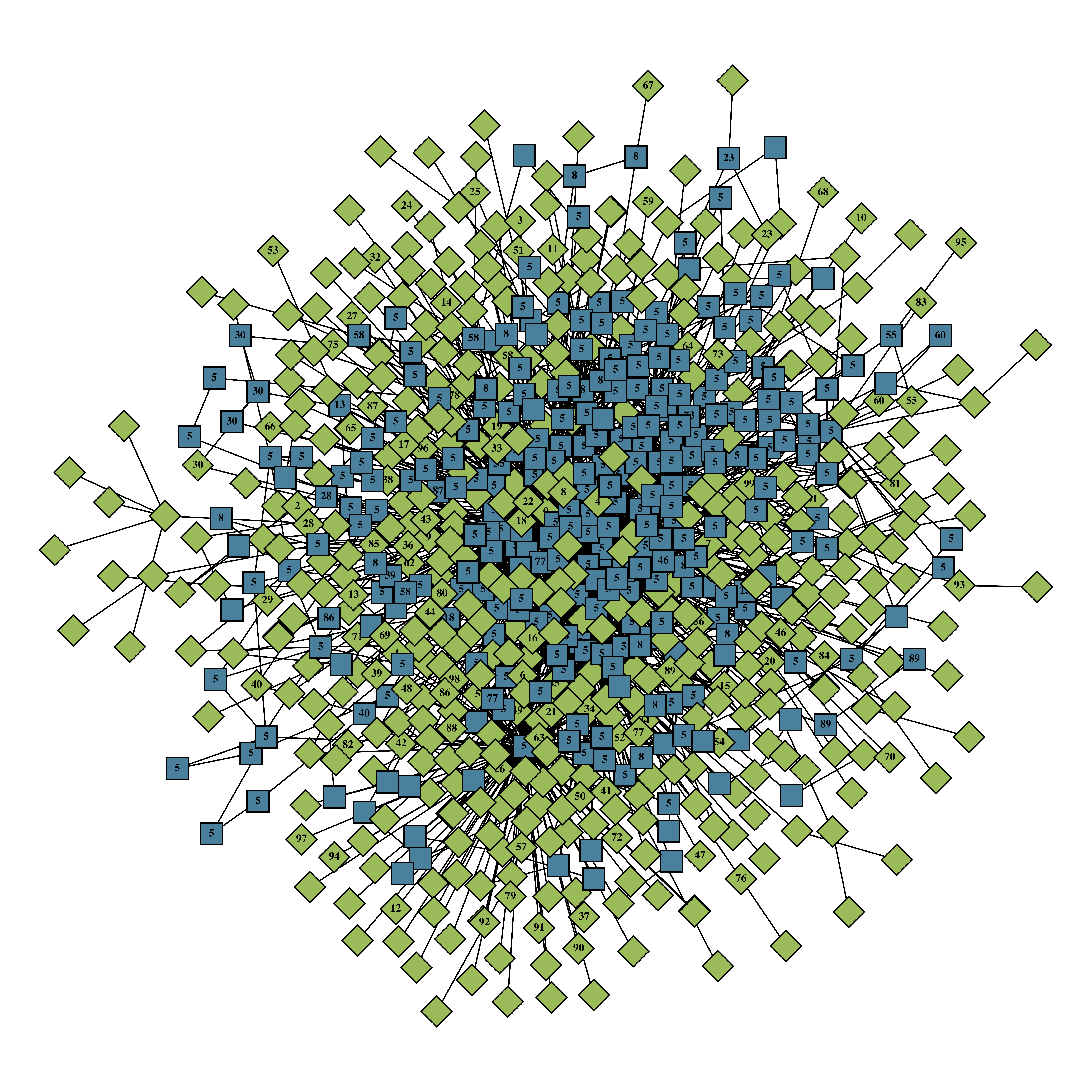}%
	\includegraphics[width=0.33\textwidth]{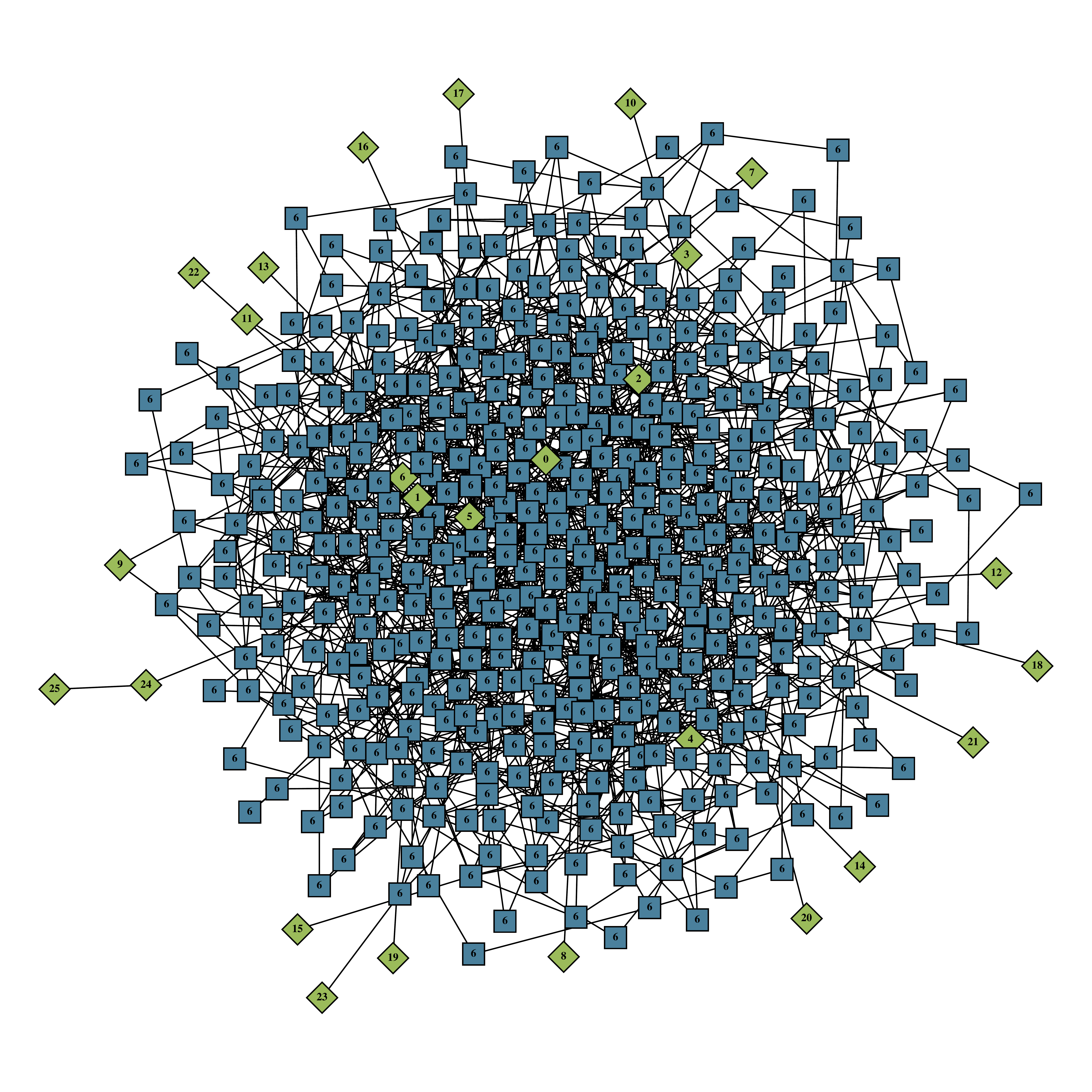}%
	\caption{\label{fig:forms}Expansion of convex subsets of nodes in (\emph{left})~globally convex collaboration graph, (\emph{middle})~regionally convex Internet map and (\emph{right})~locally convex random graph. Graphs show particular realizations of convex subsets grown from the most central node with the smallest geodesic distance $\ell$ to other nodes. The nodes included in the growing subsets by construction are shown with \dfn{diamonds}, while \dfa{squares} are the nodes included by expansion to convex subsets. The labels of the nodes indicate the step $t$ in which they were included, while the layouts were computed with Large Graph Layout~\protect\cite{ADWM04}.}
\end{figure}

\secsref{expans}{freqs} explore convexity from a local and global perspective, and demonstrate various forms of convexity in graphs and networks (see~\figref{forms}). In contrast to other networks, convex subsets expand very slowly in tree-like infrastructure networks and clique-like collaboration graph. We refer to this as \emph{global convexity}. On the other hand, random graphs are \emph{locally convex} meaning that any small connected subgraph is almost certainly convex. Finally, core-periphery networks consist of a non-convex c-core surrounded by a convex periphery, which we denote \emph{regional convexity}. Note, however, that convex periphery is only a specific type of regional convexity.

In what follows, we introduce different measures of convexity in graphs and networks. In~\secref{globals}, we propose a measure called $c$-convexity suitable for assessing global and regional convexity, while, in~\secref{locals}, we propose different measures of local convexity.

\subsection{\label{sec:globals}Global and regional convexity in networks}

Global and regional convexity in graphs and networks can be assessed by measuring the growth of convex subsets within our algorithm presented in~\secref{expans}. Recall $s(t)$ being the fraction of nodes included in the convex subsets at step $t$ of the algorithm, $s(t)\geq (t+1)/n$, where $n$ is the number of nodes in a network. Furthermore, let $t'$ be the number of steps needed for the convex subsets to expand to the entire network, $s(t')=1$. For notational convenience, let $s(t)=1$ for every $t\geq t'$.

We define the growth at step $t$ of the algorithm as $\Delta s(t)=s(t)-s(t-1)$, which is compared against the growth in an appropriate null model. A common choice is to select some random graph model in order to eliminate the effects that are merely an artifact of network density or node degree distribution~\cite{ER59,NSW01}. However, random graphs are locally convex and thus can not be used as a non-convex null model. For this reason, we rather compare networks, and also random graphs, with a fully convex graph. Such graph is a collection of cliques connected together in a tree-like manner in which the growth equals $1/n$ at each step $t$. Notice that this definition includes also a tree and a complete graph that are both convex graphs.

Let $c$ be a free parameter properly explained below, $c\geq 1$. We define $c$-convexity $X_c$ of a network as the difference $\Delta s(t)-1/n$ over all steps $t$ of the algorithm, which is subtracted from one in order to get higher values in convex networks where $\Delta s(t)\approx 1/n$. Hence,
\begin{eqnarray}
	X_c & = & 1-\sum_{t=1}^{t'}\sqrt[c]{\Delta s(t)-1/n} \nonumber \\
	& = & 1-\sum_{t=1}^{t'}\sqrt[c]{s(t)-s(t-1)-1/n} \label{eq:Xc} \\
	& = & 1-\sum_{t=1}^{n-1}\sqrt[c]{\max\{ s(t)-s(t-1)-1/n,\,0\,\}}. \label{eq:Xn}
\end{eqnarray}

For $c=1$, most of the terms of the sum in~\eqref{Xc} cancel out and $1$-convexity $X_1$ can be written as
\begin{eqnarray}
	X_1 & = & 1-\sum_{t=1}^{t'}s(t)-s(t-1)-\frac{1}{n} \nonumber \\
	& = & 1-s(t')+s(0)+\frac{t'}{n} \nonumber \\
	& = & \frac{t'+1}{n}. \label{eq:X1} 
\end{eqnarray}
$X_1$ simply measures the number of steps needed to cover the network $t'$ relative to its size $n$, $X_1\in[0,1]$. In this way, $X_1$ is an estimate of global convexity in a network. In core-periphery networks, $X_1$ can also be interpreted in terms of a non-convex c-core defined in~\secref{cores}. Let $n_c$ be the number of nodes in the network c-core and let $\avg{\ell}$ be the average distance between the nodes. Then,
\begin{eqnarray}
	X_1 & \approx & \frac{2\avg{\ell}+1}{n}+\frac{n-n_c}{n} \nonumber \\
	& = & 1-\frac{n_c-2\avg{\ell}-1}{n},
\end{eqnarray}
where $2\avg{\ell}$ is approximately the number of steps needed for the convex subsets to cover the network c-core, which actually occurs at some step $\avg{\ell}<t<2\avg{\ell}$ as shown in~\secref{sizes}, and $n-n_c$ is the number of nodes in the convex periphery. In this way, $X_1$ is an estimate of regional convexity in a network.

For $c>1$, $c$-convexity $X_c$ also takes into account the growth of convex subsets itself by emphasizing any superlinear growth in $s(t)$. Consequently, $X_c$ becomes negative in a network with a sudden expansion of convex subsets as it occurs in random graphs, $X_c\in(-\infty,1]$. Note that the sum in~\eqref{Xn} does not have to be computed entirely. After the growth of convex subsets settles, all subsequent terms of the sum are zero or close to zero and thus negligible even for very large networks. We approximate $X_c$ from the first $100$ terms of the sum in~\eqref{Xn}, which is sufficient for our purposes here.

\begin{table}%
	\caption{\label{tbl:globals}\emph{Global and regional convexity in graphs and networks. Columns show $c$-convexity of empirical networks $X_c$, randomly rewired networks $\cmd{X}_{c}$ and the corresponding Erd{\H o}s-R{\'e}nyi random graphs $\rnd{X}_{c}$. The values are estimates of the mean over $100$ runs.}}
	\begin{tabular}{lrrrrrr} \hline\hline
		Network & $X_{1}$ & $\cmd{X}_{1}$ & $\rnd{X}_{1}$  & $X_{1.1}$ & $\cmd{X}_{1.1}$ & $\rnd{X}_{1.1}$ \\\hline
		\power & $0.95$ & $0.32$ & $0.24$ & $0.91$ & $0.10$ & $0.01$ \\
		\euros & $0.66$ & $0.23$ & $0.27$ & $0.44$ & $-0.02$ & $0.06$ \\
		\collabs & $0.91$ & $0.09$ & $0.06$ & $0.83$ & $-0.05$ & $-0.09$ \\
		\oreg & $0.68$ & $0.36$ & $0.06$ & $0.53$ & $0.20$ & $-0.09$ \\
		\celeg & $0.57$ & $0.54$ & $0.07$ & $0.43$ & $0.40$ & $-0.13$ \\
		\flights & $0.43$ & $0.24$ & $0.00$ & $0.30$ & $0.16$ & $-0.07$ \\	
		\cites & $0.24$ & $0.16$ & $0.02$ & $0.04$ & $0.00$ & $-0.13$ \\
		\blogs & $0.17$ & $0.12$ & $0.00$ & $0.06$ & $0.04$ & $-0.08$ \\	
		\fweb & $0.03$ & $0.03$ & $0.02$ & $-0.06$ & $-0.02$ & $-0.02$ \\\hline\hline
	\end{tabular}
\end{table}

\tblref{globals} shows $c$-convexity $X_c$ of empirical networks from~\tblref{nets}, randomly rewired networks\revc{1}{16}{} % or the configuration model graphs 
$\cmd{X}_{c}$ and the corresponding Erd{\H o}s-R{\'e}nyi random graphs $\rnd{X}_{c}$. The results confirm our observations from~\secref{expans}. $c$-convexity is much higher in networks than random graphs with $X_c>\cmd{X}_{c}>\rnd{X}_{c}$ in all cases except the food web. This is best observed in the values of $1.1$-convexity $X_{1.1}$ that are negative in random graphs $\rnd{X}_{1.1}<0$. Standard models of small-world and scale-free networks~\cite{WS98,BA99} also fail to reproduce convexity in these networks (results not shown). According to $1$-convexity $X_{1}$, most convex networks are tree-like power grid and clique-like coauthorship graph with $X_1>0.9$. Both these networks are globally convex. On the other hand, the food web is the only network that is truly non-convex with $X_{1.1}<0$. \ces protein network represents a particular case of a regionally convex network with $X_1>0.5$ which is merely a consequence of its degree distribution $X_1\approx\cmd{X}_{1}$. Other regionally convex networks are also the Internet map and airline transportation network with $X_1\approx 0.5$. Convexity of information networks, however, is very moderate with~$X_{1.1}\approx 0$.

\revc{1}{17}{The} $c$-convexity is a global measure of convexity in graphs and networks such as the hull number of a graph~\cite{ES85} introduced in~\secref{intro}. Yet, it has a number of advantages over the hull number. It is not sensitive to small perturbations, has polynomial computational complexity and also a clear interpretation in core-periphery networks.

\subsection{\label{sec:locals}Local convexity in networks}

Local convexity in graphs and networks can be assessed either by measuring the growth of convex subsets in the first few steps of our algorithm or by computing the probabilities of convex subgraphs as done in~\secref{freqs}. We start with the latter.

As before, consider subgraphs $G_i$ with up to four nodes shown in~\figref{subgraphs}. Recall $g_i$~being the number of induced subgraphs $G_i$ in a network, $c_i$ the number of these that are convex and $P_i$ the empirical probability that a subgraph $G_i$ is convex defined in~\eqref{Pi}. The probability $P$ that a randomly selected subgraph of a network is convex is then
\begin{eqnarray}
	P & = & \sum_i \frac{g_i}{\sum_i g_i} P_i \nonumber \\
	& = & \frac{\sum_i c_i}{\sum_i g_i}. \label{eq:Pc}
\end{eqnarray}
Furthermore, let $\rnd{g}_i$ be the average number of induced subgraphs $G_i$ in a corresponding Erd{\H o}s-R{\'e}nyi random graph and $\rnd{P}_i$ the analytical probability that a subgraph $G_i$ is convex derived in~\appref{probs}. The probability $\rnd{P}$ that a randomly selected subgraph of a random graph is convex is then
\begin{eqnarray}
	\rnd{P} & = & \sum_i \frac{\rnd{g}_i}{\sum_i \rnd{g}_i} \rnd{P}_i.
\end{eqnarray}

First two columns of~\tblref{locals} show the probability of convex subgraphs $P$ in empirical networks from~\tblref{nets} and the corresponding random graphs $\rnd{P}$. Observe that the probability is much higher in random graphs than networks with $P<\rnd{P}$ in all cases except the food web. As shown in~\secref{sizes}, random graphs are locally convex with $\rnd{P}\approx 100\%$ as long as the average distance between the nodes $\ln{n}/\ln{\avg{k}}$ is larger than the size of the subgraphs. Notice that globally convex infrastructure networks are also fairly locally convex with $P\approx 80\%$. Some local convexity $P>50\%$ is observed also in regionally convex networks such as the Internet map and \ces protein network, where most of the subgraphs are found in the convex periphery. On the other hand, airline transportation network is not locally convex with $P\approx 0\%$, even though the network is regionally convex. However, as one can observe in~\figref{exmpls}, the periphery of airline transportation network consists of mostly pendant nodes, which is ignored by the subgraphs.

\begin{table}%
	\caption{\label{tbl:locals}\emph{Local convexity in graphs and networks. Columns show the probability of convex subgraphs in empirical networks $P$ and the corresponding Erd{\H o}s-R{\'e}nyi random graphs $\rnd{P}$, and the maximum size of convex subsets in networks $L_c$ and graphs $\rnd{L}_{c}$. The values in brackets are the values of $L_t$ that are different from $L_1$, while last column is the analytically derived estimate of $\rnd{L}_{1}$. The values are estimates of the mean over $100$ runs.}}
	\begin{tabular}{lrrrrr} \hline\hline
		Network & $P$ & $\rnd{P}$ & $L_1$ ($L_t$) & $\rnd{L}_1$ & $\ln{n}/\ln{\avg{k}}$ \\\hline
		\power & $77.0\%$ & $99.4\%$ & $6$ ($14$) & $9$ & $8.66$ \\
		\euros & $83.2\%$ & $97.6\%$ & $7$ ($16$) & $7$ & $7.54$ \\
		\collabs & $53.3\%$ & $71.3\%$ & $7$ ($17$) & $4$ & $3.77$ \\
		\oreg & $56.0\%$ & $86.4\%$ & $3$ & $4$ & $4.40$ \\
		\celeg & $77.8\%$ & $97.6\%$ & $2$ & $5$ & $5.79$ \\
		\flights & $5.5\%$ & $12.9\%$ & $2$ & $3$ & $2.38$ \\
		\cites & $30.5\%$ & $89.2\%$ & $3$ & $4$ & $4.30$ \\
		\blogs & $2.7\%$ & $6.0\%$ & $2$ & $2$ & $2.15$ \\
		\fweb & $2.2\%$ & $0.3\%$ & $2$ & $2$ & $1.59$ \\\hline\hline
	\end{tabular}
\end{table}

In the remaining, we assess local convexity in graphs and networks also by measuring the growth of convex subsets in the first few steps of our algorithm from~\secref{expans}. In particular, we measure the number of steps for which the convex subsets grow one node at a time and therefore no expansion occurs. The fraction of nodes $s(t)$ included in the convex subsets at step $t$ of the algorithm must thus be $s(t)\approx (t+1)/n$. This gives an estimate of local convexity seen as the maximum size of the subsets of nodes that are still expected to be convex. Note that this is different than above where we have fixed the maximum size and also the type of subgraphs, while we grow random subsets of nodes below.

We define the maximum size of convex subsets $L_c$ as
\begin{eqnarray}
	L_c & = & 1+\max\{\,t\mid s(t)<(t+c+1)/n\,\}, \label{eq:Lc}
\end{eqnarray}
where $c$ is a free parameter different than in~\eqref{Xc}, $c>0$. For $c=1$, $L_1$ measures the maximum size of the subsets of nodes that on average require less than one additional node in order to be convex $s(t)<(t+2)/n$. For $c=t$, one gets a more relaxed definition $L_t$ requiring that less than one additional node needs be included for each node in the subset or, equivalently, at each step $t$ of the algorithm $s(t)<(2t+1)/n$. To account for randomness, we use the lower bound of the $99\%$ confidence interval of $s(t)$ in~\eqref{Lc}.

Second two columns of~\tblref{locals} show the maximum size of convex subsets $L_c$ in empirical networks from~\tblref{nets} and the corresponding Erd{\H o}s-R{\'e}nyi random graphs $\rnd{L}_c$. Consider first the values of $L_1$ and $\rnd{L}_1$. These further confirm that random graphs are locally more convex than networks with $L_1\leq\rnd{L}_1$ in all cases except one. Notice also that $\rnd{L}_1$ well coincides with the analytical estimate for random graphs $\ln{n}/\ln{\avg{k}}$ derived in~\appref{sizes}. In regionally convex or non-convex networks, only very small subsets of nodes are expected to be convex with $L_1\leq 3$. On the contrary, much larger subsets are convex in globally convex infrastructure networks and collaboration graph with $L_1\approx 7$. 

Considering also the relaxed definitions $L_t$ and $\rnd{L}_t$, only three values change in~\tblref{locals}. The maximum size of convex subsets more than doubles in globally convex networks with $L_t\approx 16$, while the values remain exactly the same in all other networks $L_t=L_1$ and random graphs $\rnd{L}_t=\rnd{L}_1$. Hence, under this loose definition, globally convex networks are actually even more locally convex than random graphs.

Global convexity in networks thus implies also strong local convexity. Regionally convex networks, however, are not necessarily locally convex. This is due to a specific type of regional convexity observed in networks. Although a convex periphery can cover a large majority of the nodes in a network, these are by definition disconnected and are connected only through a non-convex c-core as shown in~\secref{cores}. Therefore, one can not grow large convex subsets solely out of the nodes in the periphery.

% % % % % % % % % % % % % % % % % % % % % % % % % % 
%
%				CONCLUSIONS
%
% % % % % % % % % % % % % % % % % % % % % % % % % %

\section{\label{sec:concs}Conclusions}

In this paper we have studied convexity in complex networks through mathematical definition of a convex subgraph. We explored convexity from a local and global perspective by observing the expansion of convex subsets of nodes and the frequency of convex subgraphs. We \reva{2}{have} demonstrated three distinct forms of convexity in graphs and networks.

Global convexity refers to a tree-like or clique-like structure of a network in which convex subsets grow very slowly and thus any connected subset of nodes is likely to be convex. Globally convex networks are spatial infrastructure networks and social collaboration graphs. This is in contrast with random graphs~\cite{ER59}, where there is a sudden expansion of convex subsets when their size exceeds $\ln{n}/\ln{\avg{k}}$ nodes. In fact, the only network studied that is globally less convex than a random graph is the food~web.

Random graphs, however, are locally convex meaning that any connected subgraph with up to $\ln{n}/\ln{\avg{k}}$ nodes is almost certainly convex. Globally convex networks are also fairly locally convex, or even more convex than random graphs under a loose definition of local convexity, whereas almost any other network studied is locally less convex than a random graph. On the other hand, most of these networks are regionally convex.

Regional convexity refers to any type of heterogeneous network structure that is only partly convex. For instance, networks with core-periphery structure can be divided into a non-convex c-core surrounded by a convex periphery. Such are the Internet map, \ces protein and airline transportation networks. Note that this type of regional convexity does not necessarily imply local convexity. This is because the nodes in convex periphery are generally disconnected and are connected only through the non-convex c-core.

We have proposed different measures of local, regional and global convexity in networks. Among them, $c$-convexity can be used to assess global convexity and measures whether the structure of a network is either tree-like or clique-like, %which is in contrast with the structure of a random graph
\revc{1}{18}{differently from random graphs}. There are many measures that separate networks from random graphs like the average node clustering coefficient~\cite{WS98} and network modularity~\cite{NG04}. However, these clearly distinguish between tree-like structure of infrastructure networks and clique-like structure of collaboration graphs. Yet, the two regimes are equivalent according to $c$-convexity. This is because they represent the border cases of networks with deterministic structure. The fact might be interesting since many of the results in network science are known only for tree-like or locally tree-like networks~\cite{DGM08,New10}. 

Convexity is thus an inherent structural property of \reva{3}{many} networks. Random graph models~\cite{ER59,NSW01} and also standard network models~\cite{WS98,BA99} fail to reproduce convexity in networks. This is not surprising as most models are based on the existence of individual edges between the nodes and not on the inclusion of the entire geodesic paths. Development of realistic models of network convexity represents an important direction for future work. The study of convexity should also be extended to directed and weighted networks, where the definition of geodesic paths changes, \revc{1}{14}{and disconnected networks, where geodesic paths are ill-defined.}

Frequency of small subgraphs is often used in network comparison or alignment~\cite{Prv07a,YMDLJKSP14} and for revealing distinct classes of networks~\cite{MIKLSASA04}. While the frequency of non-induced subgraphs or motifs~\cite{MSIKCA02} must necessarily be compared against some null model, induced subgraphs or graphlets~\cite{PCJ04} are specific local structures found in networks. Moreover, convex subgraphs are special types of induced subgraphs and might as such enable even more detailed classification of networks. This could also represent an important contribution to understanding geometric properties of networks~\cite{CE16a,CE16b}.

Another prominent direction for future work is to investigate convexity in the context of different mesoscopic structures observed in networks. We have studied convexity only in the case of core-periphery structure~\cite{BE00} and proposed a novel characterization of core and periphery, which might be of separate interest. However, other examples include also node communities~\cite{GN02}, overlapping communities~\cite{PDFV05}, disassortative groups~\cite{NL07}, role models~\cite{RW07} and hierarchical structure~\cite{CMN08}. \revc{1}{6}{In fact, many community detection methods adopt a definition of community that can actually be seen as an approximation of a convex subgraph}~\cite{RB08}.

Network convexity is an indication of uniqueness of geodesic paths in a network. The geodesic paths are mostly unique in convex infrastructure networks due to high cost of connections, while largely redundant in a non-convex food web in order for the ecosystem to survive. Convex networks thus represent locally self-sufficient systems. As such convexity can be seen as a measure of network redundancy, a concept closely related to robustness and resilience~\cite{AJB00,Bar16}.

Convexity is probably most commonly associated with its benefits in mathematical optimization~\cite{BV04a,HLB15}. For instance, many NP-complete problems in general graphs become polynomial in chordal graphs~\cite{Gav74} which can be seen as a model of convex graphs. There seems to be no good reason why such results could not be derived also for convex networks.

% % % % % % % % % % % % % % % % % % % % % % % % % % 
%
%				ACKNOWLEDGMENTS
%
% % % % % % % % % % % % % % % % % % % % % % % % % %

\section*{Acknowledgments}

The authors thank Tim Evans, Vladimir Batagelj, Jure Leskovec and Zoran Levnaji\'{c} for valuable comments and suggestions, and Ludo Waltman for providing citation data from the Web of Science bibliographic database. This work has been supported in part by the Slovenian Research Agency under programs P1-0297 and P2-0359.

% % % % % % % % % % % % % % % % % % % % % % % % % % 
%
%				BIBLIOGRAPHY
%
% % % % % % % % % % % % % % % % % % % % % % % % % %

% \bibliographystyle{notnws} % nws
% \bibliography{bibliography}

% % % % % % % % % % % % % % % % % % % % % % % % % % 
%
%				APPENDICES
%
% % % % % % % % % % % % % % % % % % % % % % % % % %

\appendix

% \section{\label{app:expans}Algorithm for expansion of convex subsets}

\section{\label{app:sizes}Size of convex subsets in random graphs}

% In this appendix, we approximate the probability $\rnd{P}_G$ that a random induced subgraph $G$ is convex in a random graph~\cite{ER59}. We then show that any connected subset with up to $\ln{n}/\ln{\avg{k}}$ nodes is almost certainly convex, where $n$ is the number of nodes in the graph, $p$ the probability of an edge and $\avg{k}$ the expected node degree, $\avg{k}=p(n-1)$.

In this appendix, we approximate the probability $\rnd{P}_G$ that a random induced \revc{1}{2}{connected} subgraph $G$ is convex in a random graph~\cite{ER59}. \hl{The sampling procedure generating subgraphs $G$ is selecting the nodes of the graph randomly with constant probability and discarding any induced subgraphs that are not connected. Thus, each induced connected subgraph $G$ with a fixed number of nodes has the same probability of being generated. Note that this procedure is not equivalent to the expansion of convex subsets of nodes in}~\secref{expans}, \hl{but we use it as an indicator of the convex growth in random graphs.}

Let $n$ be the number of nodes in a random graph, $p$ the probability of an edge and $\avg{k}$ the expected node degree, $\avg{k}=p(n-1)$. \hl{Our results below are of asymptotic nature. We say that a property holds almost certainly if the probability that it holds in a random graph with the expected node degree $\avg{k}\gg 1$ limits to one when the size of the graph $n$ grows to infinity. In particular}, we show that any \hl{induced} connected \hl{subgraph $G$} with up to $\ln{n}/\ln{\avg{k}}$ nodes is almost certainly convex.

% Since subgraph $G$ must be connected, it must include a spanning tree on $s-1$ edges. Furthermore, $G$ is a subgraph of a random graph with edge probability $p$. The probability of existence of any of ${s \choose 2}-(s-1)={s-1 \choose 2}$ possible edges of subgraph $G$ that are not on the spanning tree of $G$ is $p$. The probability that subgraph $G$ has no additional edge besides the edges of the spanning tree is then
% \begin{eqnarray}
% 	(1-p)^{s-1 \choose 2} & \approx & e^{-p{s-1 \choose 2}} \nonumber \\
% 	& = & e^{-\frac{\avg{k}(s-1)(s-2)}{2(n-1)}}. \nonumber
% \end{eqnarray}
% For $s^2\ll n$, the above expression is close to one. Thus, a random induced subgraph $G$ with much less than $\sqrt{n}$ nodes is almost certainly a tree. Since each tree on nodes $1,2,\dots,s$ has the same probability to appear, we deduce that subgraph $G$ is a random tree. The average geodesic distance and also the diameter of a random tree on $s$ nodes is almost certainly in $\cmp{\sqrt{s}}$~\cite{MM70,RS67}, while the exact expressions are $\sqrt{\frac{\pi}{2}s}$ and $2\sqrt{2\pi s}$, respectively.

\hl{We first analyze the internal structure of subgraph $G$. Denote its nodes with $1,2,\dots,s$.} Since subgraph $G$ \hl{is} connected, it must include a spanning tree on $s-1$ edges. Furthermore, $G$ is a subgraph of a random graph with edge probability $p$. The probability of existence of any of ${s \choose 2}-(s-1)={s-1 \choose 2}$ possible edges of subgraph $G$ that are not on the spanning tree of $G$ is $p$. The probability that subgraph $G$ has no additional edge besides the edges of the spanning tree is then
\begin{eqnarray}
	(1-p)^{s-1 \choose 2} & \approx & e^{-p{s-1 \choose 2}} \nonumber \\
	& = & e^{-\frac{\avg{k}(s-1)(s-2)}{2(n-1)}}. \nonumber
\end{eqnarray}
For $s^2\ll n$, the above expression is close to one. Thus, a random induced \hl{connected} subgraph $G$ with much less than $\sqrt{n}$ nodes is almost certainly a tree. Since each tree on nodes $1,2,\dots,s$ has the same probability to appear, we deduce that subgraph $G$ is a random tree. The average geodesic distance and also the diameter of a random tree on $s$ nodes is almost certainly in $\cmp{\sqrt{s}}$~\cite{MM70,RS67}, while the exact expressions are $\sqrt{\frac{\pi}{2}s}$ and $2\sqrt{2\pi s}$, respectively.

% Denote the nodes of subgraph $G$ with $1,2,\dots,s$. Notice that subgraph $G$ is convex iff no two nodes of $1,2,\dots,s$ have a geodesic path that is internally disjoint from $G$. Let $A_{ij}$ be the event that the geodesic distance between nodes $i$ and $j$ in subgraph $G$ is smaller than the geodesic distance between $i$ and $j$ in the remaining graph obtained after removing all the nodes of $G$ but $i$ and $j$, $i,j\leq s$. The event $A_{ij}$ occurs iff nodes $i$ and $j$ have no geodesic path completely outside of subgraph $G$. Hence, the probability $\rnd{P}_G$ equals
% \begin{eqnarray}
% 	\rnd{P}_G & = & \prob{\cap_{1\leq i<j\leq s}\, A_{ij}}. \nonumber
% \end{eqnarray}
% For $s\ll n$, the events $A_{ij}$ have a small correlation if any, thus
% \begin{eqnarray}
% 	\rnd{P}_G & \approx & \prod_{1\leq i<j\leq s} \prob{A_{ij}}. \nonumber
% \end{eqnarray}
% Moreover, the remaining graph is approximately a random graph on $n-s+2\approx n$ nodes with the same edge probability $p$, except for the edge between nodes $i$ and $j$ which is fixed by construction. Let $d_{ij}$ be the geodesic distance between nodes $i$ and $j$ in subgraph $G$, $i,j\leq s$. The probability that the geodesic distance between $i$ and $j$ in the remaining graph is at most $d_{ij}$ is $e^{-\frac{\avg{k}^{d_{ij}}}{n}}$~\cite{New10}. Hence,
% \begin{eqnarray}
% 	\rnd{P}_G & \approx & \prod_{1\leq i<j\leq s} e^{-\frac{\avg{k}^{d_{ij}}}{n}}. \nonumber
% \end{eqnarray}

Notice that subgraph $G$ is convex iff no two nodes of $1,2,\dots,s$ have a geodesic path that is internally disjoint from $G$. \hl{Besides, for each node $i$ of subgraph $G$, the remaining graph obtained after removing all the nodes of $G$ but $i$ is still a random graph with edge probability $p=\avg{k}/(n-1)$ and the average degree $\avg{k'}=p(n-s)$. In such a graph, the expected number of nodes at the geodesic distance at most $d$ from node $i$ is approximately $1+\avg{k'}+\avg{k'}(\avg{k'}-1)+\dots+\avg{k'}(\avg{k'}-1)^{d-1}\leq\avg{k'}^d\leq\avg{k}^d$}~\cite{New10}. \hl{We denote with $\ball{i}{d}$ the set of nodes at the geodesic distance at most $d$ from a given node $i$ also called the $d$th neighborhood of node $i$, $i\leq s$.}

% By the above results, the diameter of subgraph $G$ is almost certainly $c_1\sqrt{s}$ for some constant $c_1>0$. Observing that the geodesic distance between any two nodes in subgraph $G$ is then at most $c_1\sqrt{s}$, we obtain
% \begin{eqnarray}
% 	\rnd{P}_G & \approx & \prod_{1\leq i<j\leq s} e^{-\frac{\avg{k}^{d_{ij}}}{n}} \nonumber \\
% 	& \geq & e^{-\frac{\avg{k}^{c_1\sqrt{s}}}{n}{s \choose 2}}. \nonumber 
% \end{eqnarray}
% For $s \leq \frac{1}{c_1^2}\log_{\avg{k}}{n}$ and large enough $n$, the probability $\rnd{P}_G$ is close to one meaning that subgraph $G$ is convex. 

\hl{Let $r$ be a number such that $2r+1$ is greater than the diameter of subgraph $G$ but as close as possible. By the above results, $2r+1\approx c_1\sqrt{s}+\cmp{1}$ for some constant $c_1>0$. If for each pair of nodes $i$ and $j$ of subgraph $G$ there is no edge connecting the nodes in the $r$th neighborhood $\ball{i}{r}$ with the nodes in the $r$th neighborhood $\ball{j}{r}$, subgraph $G$ must be convex, $i,j\leq s$. Since the sizes of these neighborhoods are on average smaller than $\avg{k}^r$, there is at most $\avg{k}^r\avg{k}^r{s \choose 2}$ possible edges. Hence, the probability $\rnd{P}_G$ that subgraph $G$ is convex is}
\begin{eqnarray}
	\rnd{P}_G & \geq & (1-p)^{\avg{k}^{2r}{s \choose 2}}. \nonumber
\end{eqnarray}
\hl{Taking the logarithm of both sides we find}
\begin{eqnarray}
	\ln{\rnd{P}_G} & \geq & \avg{k}^{2r}{s \choose 2}\ln\left(1-\frac{\avg{k}}{n-1}\right) \nonumber \\ 
	& \approx & \frac{\avg{k}^{2r+1}}{n-1}{s \choose 2}, \nonumber
\end{eqnarray}
\hl{and thus}
\begin{eqnarray}
	\rnd{P}_G & \geq & e^{-\frac{\avg{k}^{c_1\sqrt{s}+\cmp{1}}}{n-1}{s \choose 2}}. \nonumber
\end{eqnarray}
For \hl{$s\leq\log_{\avg{k}}{n}$} and large enough $n$, the probability $\rnd{P}_G$ is close to one meaning that subgraph $G$ is convex. 

% On the other hand, since the average geodesic distance in subgraph $G$ is almost certainly $c_2\sqrt{s}$ for some constant $c_2>0$, there is a non-trivial fraction $f>0$ of nodes at the geodesic distance at least $c_2\sqrt{s}$. Hence,
% \begin{eqnarray}
% 	\rnd{P}_G & \approx & \prod_{1\leq i<j\leq s} e^{-\frac{\avg{k}^{d_{ij}}}{n}} \nonumber \\
% 	& \leq & e^{-\frac{\avg{k}^{c_2\sqrt{s}}}{n}f{s \choose 2}}. \nonumber 
% \end{eqnarray}
% For $s \geq \frac{1}{c_2^2}\log^2_{\avg{k}}{n}$ and large enough $n$, the probability $\rnd{P}_G$ is close to zero meaning that subgraph $G$ is not convex.

On the other hand, since the average geodesic distance in subgraph $G$ is almost certainly $c_2\sqrt{s}$ for some constant $c_2>0$, there is a non-trivial fraction $f>0$ of nodes at the geodesic distance at least $c_2\sqrt{s}$. \hl{Similar as before let $r$ be a number such that $2r+1$ is less then the average geodesic distance in subgraph $G$ but as close as possible, $2r+1\approx c_2\sqrt{s}-\cmp{1}$. Recall that the expected number of nodes in the $r$th neighborhood $\ball{i}{r}$ at the geodesic distance exactly $r$ from node $i$ is approximately $\avg{k'}(\avg{k'}-1)^{r-1}$}~\cite{New10} \hl{which is greater than $(\avg{k}/2)^r$ assuming $\avg{k}\gg 1$ and $s^2\ll n$. These nodes are also called the surface of the neighborhood. Now let $i$ and $j$ be a pair of nodes of subgraph $G$ at the geodesic distance at least $2r+1$ in $G$, $i,j\leq s$. If the $r$th neighborhoods $\ball{i}{r}$ and $\ball{j}{r}$ are not disjoint, then subgraph $G$ is not convex. Assuming that the neighborhoods are disjoint, subgraph $G$ is still not convex if there exists an edge between the nodes of the surfaces of the two neighborhoods. Let $A$ be the event that all ${s \choose 2}$ pairs of $r$th neighborhoods $\ball{i}{r}$ and $\ball{j}{r}$ are disjoint, and let $A_G$ be the event that subgraph $G$ is convex. The probability $\rnd{P}_G$ can then be written as}
\begin{eqnarray}
	\rnd{P}_G & = &  P(\given{A_G}A)P(A) \nonumber \\ 
	& \leq &  P(\given{A_G}A) \nonumber \\ 
	& \leq & (1-p)^{(\avg{k}/2)^{2r}f{s \choose 2}} \nonumber
\end{eqnarray}
\hl{and by the same sequence of arguments as above we obtain}
\begin{eqnarray}
	\rnd{P}_G & \leq & e^{-\frac{(\avg{k}/2)^{c_2\sqrt{s}-\cmp{1}}}{n-1}f{s \choose 2}}. \nonumber
\end{eqnarray}
For \hl{$s\geq\log^2_{\avg{k}/2}{n}$} and large enough $n$, the probability $\rnd{P}_G$ is close to zero meaning that subgraph $G$ is not convex.

\hl{To gain mathematical completeness of the above results one would have to analyze also non-average cases of the properties considered. In particular, one would have to analyze what if the diameter and the average geodesic distance in subgraph $G$ are not $c\sqrt{s}$ for some constant $c>0$, and what if the sizes of the $d$th neighborhoods $\ball{i}{d}$ are not $\avg{k}^d$, $i\leq s$. Since the tails of the probability distributions of these events are thin}~\cite{RS67,MM70,New10}, \hl{while we are here only interested in the asymptotic behavior and not in the exact constants of the threshold functions, we leave such analyses to be performed elsewhere.}

% Consider now the expansion of convex subsets of nodes within our algorithm from~\secref{expans}. In the first few steps of the algorithm the subsets grow one node at a time and the induced subgraphs are almost certainly convex trees. In fact, any connected subset with up to $\ln{n}/\ln{\avg{k}}$ nodes is almost certainly convex. The sudden expansion of convex subsets occurs between $\cmp{\log_{\avg{k}}{n}}$ and $\cmp{\log^2_{\avg{k}}{n}}$ nodes, while the exact threshold function is the solution $s$ of the equation $n=\avg{k}^{\sqrt{s}}{s \choose 2} $.

\hl{Finally}, consider the expansion of convex subsets of nodes within our algorithm from~\secref{expans}. In the first few steps of the algorithm the subsets grow one node at a time and the induced subgraphs are convex trees. \hl{This can be anticipated since any random induced connected subgraph with much less than $\sqrt{n}$ nodes is almost certainly a random tree and any such tree with up to $\ln{n}/\ln{\avg{k}}$ nodes is almost certainly convex.} The sudden expansion of convex subsets occurs between $\cmp{\log_{\avg{k}}{n}}$ and \hl{$\cmp{\log^2_{\avg{k}/2}{n}}$} nodes, while the exact threshold function \hl{suggested by the above calculations} is the solution $s$ of the equation $n=\avg{k}^{\sqrt{s}}{s \choose 2}$. % TODO

\section{\label{app:probs}Probability of convex subgraphs in random graphs}

\figref{subgraphs} shows all connected non-isomorphic subgraphs $G_i$ with up to four nodes. In this appendix, we derive the probabilities $\rnd{P}_i$ that a randomly selected induced subgraph $G_i$ is convex in a random graph~\cite{ER59}. As before, let $n$ be the number of nodes in the graph, $p$ the probability of an edge and $\avg{k}$ the expected node degree, $\avg{k}=p(n-1)$.

First, recall that the clique subgraphs $G_0$, $G_2$ and $G_8$ are convex by construction. 
\begin{eqnarray}
	\rnd{P}_0 & = & 1 \nonumber \\
	\rnd{P}_2 & = & 1 \nonumber \\
	\rnd{P}_8 & = & 1 \nonumber
\end{eqnarray}

Next, we consider the star subgraph $G_4$. Denote the central node of subgraph $G_4$ with $1$, the pendant nodes with $2,3,4$ and the remaining nodes with $5,6,\dots,n$. Notice that subgraph $G_4$ is convex iff no two nodes of $2,3,4$ have a common neighbor other than $1$. Let $A_i$ be the event that subgraph $G_4$ is convex in a graph induced by the nodes of $G_4\cup\set{i}$, $i\geq 5$. The event $A_i$ occurs iff node $i$ is connected to at most one of the nodes $2,3,4$. Hence,
\begin{eqnarray}
	\prob{A_i} & = & (1-p)^3+3p(1-p)^2 \nonumber \\
	& = & 1-3p^2+2p^3. \nonumber
\end{eqnarray}
Since the events $A_i$ are independent, the probability $\rnd{P}_4$ equals
\begin{eqnarray}
	\rnd{P}_4 & = & (1-3p^2+2p^3)^{n-4}. \nonumber
\end{eqnarray}

For other subgraphs with diameter two $G_1$, $G_5$, $G_6$ and $G_7$, the derivation is analogous. 
\begin{eqnarray}
	\rnd{P}_1 & = & (1-p^2)^{n-3} \nonumber \\
	\rnd{P}_5 & = & (1-2p^2+p^4)^{n-4} \nonumber \\
	\rnd{P}_6 & = & (1-2p^2+3p^3)^{n-4} \nonumber \\
	\rnd{P}_7 & = & (1-p^2)^{n-4} \nonumber
\end{eqnarray}

For the path subgraph $G_3$, one must take a different approach. Denote the nodes of subgraph $G_3$ with $1,2,3,4$, where $1,4$ are the pendant nodes. Let $A_1$ be the event that nodes $1,4$ have no common neighbor, and that nodes $1,3$ and $2,4$ have no common neighbor other than $2$ and $3$, respectively. Furthermore, let $A_2$ be the event that there is no path of length three connecting nodes $1,4$ that is outside of subgraph $G_3$. Then, $\rnd{P}_3=\prob{A_1}\prob{\given{A_2}A_1}$.

The event $A_1$ occurs iff each node $i\geq 5$ is either connected to at most one of the nodes $1,2,3,4$ or is connected to a pair of connected nodes $1,2$ or $2,3$ or $3,4$. Hence,
\begin{eqnarray}
	\prob{A_1} & = & \left((1-p)^4+4(1-p)^3+3(1-p)^2p^2\right)^{n-4} \nonumber \\
	& = & (4p^4-6p^3-3p^2+8p-3)^{n-4}. \nonumber
\end{eqnarray}
On the other hand, the probability $\prob{\given{A_2}A_1}$ can be computed as follows. The event $A_2$ occurs iff no pair of neighbors of nodes $1,4$ other than $2,3$ is connected. Since nodes $1,4$ have $c=p(n-4)$ other neighbors on average, there are $c^2$ of such pairs given $A_1$ and 
\begin{eqnarray}
	\prob{\given{A_2}A_1} & = & (1-p)^{c^2}. \nonumber
\end{eqnarray}

The probability $\rnd{P}_3$ thus reads
\begin{eqnarray}
	\rnd{P}_3 & = & (4p^4-6p^3-3p^2+8p-3)^{n-4}(1-p)^{p^2(n-4)^2}. \nonumber
\end{eqnarray}

\end{document}